\documentclass[%
 reprint,
superscriptaddress,
 amsmath,amssymb,
 aps,
]{revtex4-2}

\usepackage{graphicx}
\usepackage{dcolumn}
\usepackage{bm}
\usepackage{hyperref}

\begin{document}

\preprint{APS/123-QED}

\title{Multiple-order differential imaging based on two types of topological singularity in one dimensional photonic crystals}

\author{Haoran Zhang}
 \affiliation{Department of Illuminating Engineering and Light Sources, College of Intelligent Robotics and Advanced Manufacturing, Fudan University, Shanghai 200433, China.}
\author{Yufu Liu}%
\affiliation{Department of Illuminating Engineering and Light Sources, College of Intelligent Robotics and Advanced Manufacturing, Fudan University, Shanghai 200433, China.}
\author{Yunlin Li}%
\affiliation{Department of Illuminating Engineering and Light Sources, College of Intelligent Robotics and Advanced Manufacturing, Fudan University, Shanghai 200433, China.}
\author{Xunya Jiang}
\affiliation{Department of Illuminating Engineering and Light Sources, College of Intelligent Robotics and Advanced Manufacturing, Fudan University, Shanghai 200433, China.}
\date{\today}

\begin{abstract}
The differential imaging have garnered significant attention owing to its boundary detection capabilities in image processing. However, to date, there has been scant research investigating the relationship between the differential imaging effect and the topological properties of a one dimensional(1D) system. In this work, we systematically investigate the multiple-order differential imaging based on two types of topological singularity in 1D photonic crystals(PhCs). For both oblique and normal incidences , We conduct a detailed investigation of differential imaging effect. For the oblique incident cases, the first type topological singularities support first-order differential imaging in both x and y directions. Meanwhile, based on the second type topological singularities, $\partial^2 /\partial x \partial y$-type differential imaging can be achieved. For the normal incident cases, the first type topological singularities can support radial second-order and fourth-order differential imaging and the second type topological singularities can support radial fourth-order differential imaging without the need for fine tuning of the structural parameters. We further demonstrates the realization of these differential imaging effects in the deep subwavelength region by shifting the first type topological singularities into this region. This research connects the topological properties of PhCs with optical differential imaging, paving the way for the development of multiple-order differential imaging devices with improved robustness and functionality.

\
\end{abstract}

\maketitle


\section{\label{sec:level1}INTRODUCTION}

Over the past decade, optical computing technology, with its exceptional speed, low power characteristics, and capability for large-scale parallel processing, is widely regarded as an effective alternative to integrated circuits \cite{li2021challenges,wang2022single,fu2022ultracompact}. Particularly, spatial optical differential imaging technology has attracted significant attention due to its boundary detection capabilities in image processing \cite{wan2021review} . This technology enables real-time data compression and object classification by capturing boundary information in images, demonstrating substantial application potential in fields such as autonomous driving and biomedicine \cite{tang2000mri,rajab2004application}.

To achieve spatial optical differential imaging, numerous methods have been proposed. For instance, differential imaging at air-glass interfaces based on the spin Hall effect of light is achieved by restricting the polarization direction of incident and reflected light to be orthogonal \cite{zhu2019generalized}. Utilizing Brewster's effect and cross-polarization at optical surfaces \cite{xu2020optical,xu2021enhanced,xia2021tunable,mi2020tunable,xu2020goos,deng2024r} or the spin Hall effect of light in Weyl semimetals \cite{wen2024tunable}     enables tunable first-order differential imaging; To achieve isotropic two-dimensional differential imaging, precise adjustment of the band dispersion in photonic crystal slabs can enable effective two-dimensional Laplacian operations \cite{guo2018photonic}. Lately, Zhu et al. \cite{zhu2021topological} propose a topological optical differential device based on a single unpatterned dielectric interface under total internal reflection, the transfer function carries a topological charge of ±1, this method can achieve two-dimensional, isotropic, first order differential imaging. Up to now, most planar optical elements that achieve differential imaging effects are limited to low orders, such as first or second order \cite{zhou2020flat,zhang2017all,pan2021laplace,deng2024broadband}. Recently, high-order differential operations based on 1D PhCs have been proposed, enabling high-order differential imaging at different frequency points, such as third and fourth order, through fine-tuning of structural parameters\cite{liu2022single}.

On the other hand, the concept of nontrivial topology of band-gap has been extensively studied, and there is a growing interest in exploiting topological properties in various physical systems    \cite{hasan2010colloquium,shen2012topological,lu2014topological,xiao2014surface,bansil2016colloquium,chiu2016classification,khanikaev2017two,wang2020topological,xiong2021resonance}. Due to the fact that topological invariants of a system cannot be continuously changed under adiabatic approximation, many physical processes associated with topological properties are robust against external perturbations, such as the topological edge states \cite{asboth2013bulk,batra2019understanding}. Recently, the topological band structure of 1D PhCs has been widely investigated. Numerous studies have focused on topological singularities in the band structure, which are defined as the positions of topological charge in k-space. At these points, the phase of the Bloch modes undergoes a $\pi$ Zak phase jump, rendering the band-gap structure of 1D PhCs nontriviality \cite{xiao2014surface,xiong2021resonance,li2019two,li2019singularity,liu2023evolution}. Interestingly, the frequency of topological singularities can be directly determined through the zero-scattering condition of 1D PhCs, implying perfect transmission points within the periodic structure. Furthermore, Xiong et al. \cite{xiong2021resonance} has successfully shifted the frequency of topological singularities to the deep subwavelength region close to zero frequency using an ABCBA-kind PhCs. The deep-subwavelength is defined as $\lambda \geq20a$.  Within this frequency regime, the effective medium theory is broken down with the introduce of topological singularity due to their global topological properties \cite{liudual}.

However, despite significant efforts that have been applied to achieve optical differential imaging, the optical differential imaging induced around topological singularities in 1D PhCs has not been systematically studied. The mechanism of optical differential imaging around singularities remains unveiled. Several fundamental questions still await answers, for example, "Can we achieve first-order differential imaging in both $x-$and $y-$ directions for p-polarized and s-polarized light incidence without the restriction of Brewster's angle?" and "Can second-order or even fourth-order differential imaging be realized with topological singularities?" Furthermore, "Can the corresponding differential imaging effects be achieved in the deep subwavelength region after the singularities are introduced?" If we can discover the correlation between topological singularities and optical differential imaging, it would be of theoretical significance and facilitate the design of related devices.

\begin{figure}[htb]
	\includegraphics[width=1\linewidth]{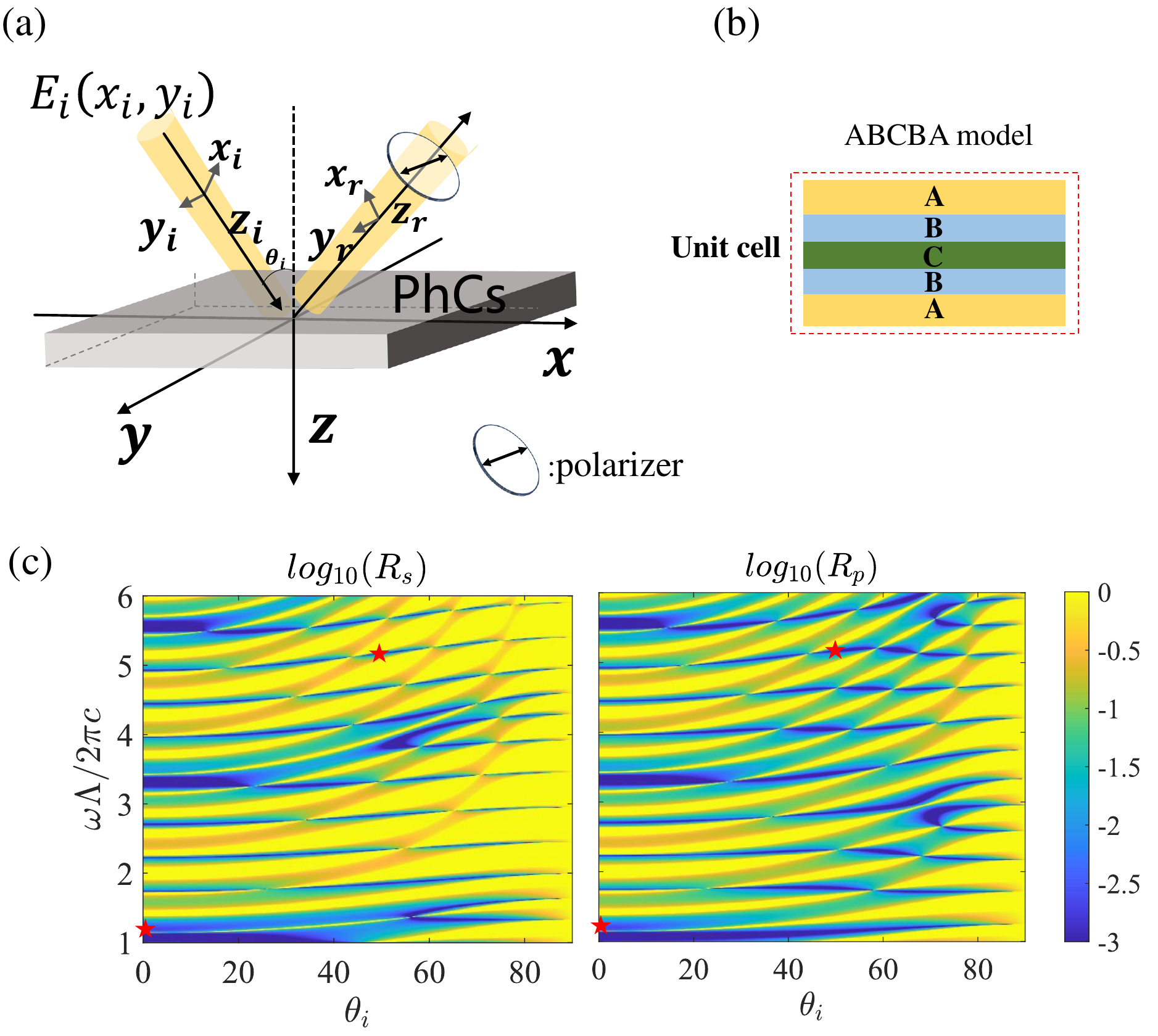}
	\caption{\label{fig:model}(a)The schematic of differential imaging based on PhCs. The polarizer are oriented at the angles indicated with double-head arrows. Here, the coordinates $x$ and $y$  are defined in the reference planes for incident and reflected beams respectively. (b)The schematic of a unit cell PhCs in x-z view which are composed of five layers (ABCBA model). (c)The reflectivity in dB of s-polarization(left panel) and p-polarization(right panel) in $\{\theta_i,\omega\}$space for semi-infinite PhCs. The red pentagon represent the second type topological singularity we used in this paper. Structural parameters are set as $d_A=0.4\Lambda$, $d_B=0.3\Lambda$, $d_C=0.3\Lambda$, $n_A=1$, $n_B=1.5$, $n_C=3$. }
\end{figure}

In this work, we systematically investigate the theory and effects of differential imaging based on topological singularities in 1D ABCBA-kind PhCs. We find around different types of topological singularities under different incident angles, multiple-order differential imaging effect can be achieved. Specifically, for the oblique incident cases, around the first type topological singularity, first-order differential imaging in $x-$ and $y-$ direction can be realized. Around the second type topological singularity, $\partial^2 /\partial x \partial y$- type differential imaging can be realized. For the normal incident cases, we find that around the first type topological singularity, radial second-order differential imaging can be realized. Furthermore, when we tune the structural or material parameter, radial fourth-order differential imaging can be realized. We also find around the second type topological singularity, radial fourth-order differential imaging can be realized without fine tuning. Based on the property that the first type of topological singularities can be moved to the near-zero frequency deep subwavelength region, the differential imaging effect mentioned above can also be achieved in the deep subwavelength region. This work builds the bridge between the study of topology in 1D PhCs and optical differential imaging. Theoretically, we have established a multiple-order differential imaging theory based on the topological singularity of 1D PhCs. From a practical perspective, our 1D model is easy to fabricate and offers advantages such as strong robustness and low absorption. Therefore, our work offers key insights for the design and fabrication of high-order differential imaging devices.
\section{Differential imaging model by topological singularities of 1D PhCs}
Our model of differential imaging is schematically shown in Fig. ~\ref{fig:model}(a). A linearly polarized beam with certain frequency is obliquely or normally incident on the surface of the 1D PhC slab with $N$ cells. When two conditions are satisfied, the reflected beam, after passing a polarizer at certain direction, could present the differential imaging of different orders. The first condition is the large beam width condition(LBWC) which means the beam width is much larger than the wavelength. So the paraxial limit is fit for our model. The second condition is the near-singularity condition(NSC), which means the frequency $\omega$ and the parallel component of the central wavevector $\boldsymbol k_{0\parallel}$ of incident beam are same as the frequency $\omega_s$ and the parallel component of the Bloch wavevector $\boldsymbol k_{s\parallel}$ of a certain topological singularity of PhC.

As shown in Fig.~\ref{fig:model}(a), we study the incident beam on the PhC which satisfies both LBWC and NSC. We define $(x,y,z)$ and $(x_{\alpha},y_{\alpha},z_{\alpha})$ as the laboratory and local coordinates, where the subscripts $\alpha=\{i,r\}$ represent the incident and reflected beams, respectively. We set $y_{\alpha}$ parallel to $y$ of laboratory coordinate and $z_{\alpha}$ parallel to the central wave vectors of the beams. Here we note that the subscripts $\alpha=\{i,r\}$ of local coordinates are neglected generally in this work for the incident and reflected beams, since they are always studied in different local coordinates.
The fields of incident and reflect beams have the form $\boldsymbol{e}_{i} E_{i}(x,y)$, and $\boldsymbol{e}_{r} E_{r}(x,y)$, where the 2-vector $\boldsymbol{e}_{{i}}=(e_i^{x},e_i^{y})^T$ and $\boldsymbol{e}_{{r}}=(e_r^{x},e_r^{y})^T$ are the normalized unit vectors for incident and reflected polarizations in their local $x$-$y$ planes, $E_{i}(x,y)$ and $E_{r}(x,y)$ are the scalar electric field distribution on the plane perpendicular to the beam propagation direction. Because the reflected beam needs to pass a polarizer, the polarization $\boldsymbol{e}_{r}$ of the reflect beam can be selected by the direction of polarizer. The field of incident beam can be decomposed into plane-wave components around its central wavevector $\boldsymbol{k_0}$ as:
\begin{equation}\label{Fourier}
	E_{i}(x,y)=\int\int \widetilde{E}_{i}(k_x,k_y)exp(ik_x x)exp(ik_y y)dk_xdk_y,
\end{equation}

Similarly, the field of the reflected beam can decomposed around its central wavevector as:
\begin{equation}\label{Fourier2}
	E_{r}(x,y)=\int\int \widetilde{E}_{r}(k_x,k_y)exp(ik_x x)exp(ik_y y)dk_xdk_y.
\end{equation}
Since the paraxial limit, both $k_x$ and $k_y$ are supposed to be small values compared with $k_0$. The relationship between the reflected field and the incident field can be described as: 
\begin{equation}\label{reflect2}
{\widetilde{E}}_r(k_x,k_y)=\boldsymbol{e}_{r}^\dagger  \boldsymbol{M}^{\dagger}\boldsymbol{{R}}(k_x,k_y)\boldsymbol{M} \boldsymbol{e}_{i} \widetilde{E}_{i}(k_x,k_y),
\end{equation}
here, the matrix $\boldsymbol{R}(k_x,k_y)$ is:
\begin{equation}\label{eq16}
	\boldsymbol{{R}}(k_x,k_y)=\left[
	\begin{array}{cc}
		r_p(k_x,k_y)& 0\\
		0&	r_s(k_x,k_y)
	\end{array}
	\right],
\end{equation}
where $r_p$ and $r_s$ are  reflection coefficients for p- and s-polarized plane waves with the incident wavevector ($k_x$, $k_y$), respectively. The matrix $\boldsymbol{M}$ in Eq. (\ref{reflect2}) is the transformation matrix that converting the wavevector from a non-central wavevector to the central wavevector.

The cell of 1D ABCBA-kind PhC is depicted in Fig. ~\ref{fig:model}(b), which consists five layers in the unit cell with refractive index as $n_A$, $n_B$, $n_C$, $n_B$, $n_A$ and the widths of five layers are set as $d_A/2$, $d_B/2$, $d_C$, $d_B/2$, $d_A/2$, respectively. The relative permeability of all dielectric materials is supposed to be $\mu_r=1$. Obviously, there are two center for spatial inversion symmetry which are the central point of layer-C and the start point of layer-A. We note that 1D ABCBA-kind PhC can be degenerated into common 1D ABA-kind PhC if we assume $d_B$ or $d_C=0$. 

Next, we will introduce the topological singularity of 1D ABCBA-kind PhCs. A topological singularity is generally defined as the zero-scattering point on the bands for 1D PhCs, where the topological charge is located\cite{xiao2014surface,liudual}.  At the singularity, we have the reflection coefficients $r_{s}=0$ for s-polarized singularities and $r_{p}=0$ for p-polarized singularities. The detailed derivation can be found in Appendix \ref{zero}.

For ABCBA-kind PhCs, there are two types of topological singularities\cite{li2019singularity}.  The first type topological singularity satisfies:
\begin{equation}\label{sgc}
	\begin{aligned}
		&F_1 \sin(k_{Cz}d_C) \frac{\cos^2(k_{Bz}d_B)}{2} - F_2 \sin(k_{Cz}d_C) \frac{\sin^2(k_{Bz}d_B)}{2} \\
		&+ F_3 \cos(k_{Cz}d_C) \sin(k_{Bz}d_B)=0, and\\ 
		\\
		& sin(k_{Bz}d_B)\neq0,  sin(k_{Cz}d_C)\neq0,
	\end{aligned}
\end{equation}
 where $k_{Az}=\sqrt{k_0^2n_A^2-k_\rho^2}$, $k_{Bz}=\sqrt{k_0^2n_B^2-k_\rho^2}$ and $k_{Cz}=\sqrt{k_0^2n_C^2-k_\rho^2}$ are the wave vectors in the z-direction within layer A, B and C, respectively. Here $k_\rho=\sqrt{(k_x^2+k_y^2)}$. $F_1$, $F_2$ and $F_3$ are the parameters, $F_1=\frac{k_{Cz}}{k_{Az}}-\frac{k_{Az}}{k_{Cz}}$, $F_2=\frac{{k_{Bz}}^2}{k_{Az}k_{Cz}}-\frac{k_{Az}k_{Cz}}{{k_{Bz}}^2}$, $F_3=\frac{k_{Bz}}{k_{Az}}-\frac{k_{Az}}{k_{Bz}}$ for $s$-wave and $F_1 = \frac{\varepsilon_C k_{Az}}{\varepsilon_A k_{Cz}} - \frac{\varepsilon_A k_{Cz}}{\varepsilon_C k_{Az}}$, $F_2=\frac{{\varepsilon_B}^2k_{Az}k_{Cz}}{\varepsilon_A\varepsilon_C{k_{Bz}}^2}-\frac{\varepsilon_A\varepsilon_C{k_{Bz}}^2}{{\varepsilon_B}^2k_{Az}k_{Cz}}$, $F_3=\frac{\varepsilon_Bk_{Az}}{\varepsilon_Ak_{Bz}}-\frac{\varepsilon_Ak_{Bz}}{\varepsilon_Bk_{Az}}$for $p$-wave.
The second type topological singularity satisfies:
\begin{equation}\label{second type}
	sin(k_{Bz}d_B)=sin(k_{Cz}d_C)=0.
\end{equation}
 We note that, when $d_C=0$ is zero, the second type topological singularity for ABCBA-kind PhC is degenerated into the well known topological singularity of ABA-kind PhC with $sin(k_{Bz}d_B)=0$ \cite{xiao2014surface}.

For the better understanding of topological singularities, we then take an example of ABCBA-kind PhC. Here we set $d_A=0.4\Lambda$, $d_B=0.3\Lambda$, $d_C=0.3\Lambda$, $n_A=1$, $n_B=1.5$, $n_C=3$, where $\Lambda$ is the length of the unit cell.  Fig.~\ref{fig:model}(c) shows the reflectivity in $dB$ of s-polarization(left panel) and p-polarization(right panel) in $\{\theta_i,\omega\}$ space for semi-infinite PhCs. First, the deep yellow regions represent the photonic gaps. Second, the deep blue lines with $|R_{s(p)}|\approx 0$ represent the evolving trajectories of first type topological singularities.  From the Fig.~\ref{fig:model}(c), we can see that there are the multiple channels of the first type topological singularities. 
At last, we note that the second type topological singularities are on discrete points in $\{\theta_i,\omega\}$ space since there are two conditions needs to satisfied in Eq. (\ref{second type}). The red pentagons in Fig.~\ref{fig:model}(c) represent two examples of second type topological singularities which will be studied in detail.


\section{DIFFERENT-ORDER DIFFERENTIAL IMAGING FROM TOPOLOGICAL SINGULARITIES OF PHCs UNDER OBLIQUE INCIDENCE}
In this section, we will show that when a linearly polarized beam which satisfies both LBWC and NSC is \textbf{obliquely} incident on the surface of the ABCBA-kind PhC and the reflected beam, after passing a polarizer, could present first-order and second-order differential imaging. In details, for the first type topological singularity, first-order optical differential imaging in both the $x-$ and $y-$ directions can be achieved. For the second type topological singularity, second-order $\partial^2 /\partial x \partial y$-type differential imaging effects can be achieved. Furthermore, we propose a deep subwavelength first-order differential imaging design, which breaks the limitations of the effective medium theory. 

\subsection {The transfer function of the system under oblique incidence}


 In Eq. (\ref{reflect2}), $\boldsymbol{M}$ is the matrix that converting the wavevector from a non-central wavevector to the central wavevector, which can be expressed in the paraxial limit as: 
\begin{equation}\label{m}
	\boldsymbol{M}=\left[
	\begin{array}{cc}
		1& \frac{k_ycot(\theta_i)}{k_0}\\
		-\frac{k_ycot(\theta_i)}{k_0}& 1
	\end{array}
	\right].
\end{equation}

Then, we can define the optical transfer function\cite{zhu2021topological} which relates the incident and reflected fields as:
\begin{equation}\label{Er}
 \widetilde{E}_{r}(k_x,k_y)=H(k_x,k_y)\widetilde{E}_{i}(k_x,k_y).
\end{equation}
According to Eq. (\ref{reflect2}-\ref{eq16}) and  Eq. (\ref{m}-\ref{Er}), the transfer function can be expressed as:
\begin{equation}\label{H}
	H(k_x,k_y)=\boldsymbol{e}_{r}^\dagger \boldsymbol{M}^{\dagger}  \boldsymbol{{R}}(k_x,k_y) \boldsymbol{M} \boldsymbol{e}_{i}.
\end{equation}

To achieve the differential imaging, the first step is to analyze the optical transfer function $H(k_x,k_y)$ in small $k_x$ and $k_y$ limit. In Eq. ({\ref{H}}), by further expanding $r_p$ and $r_s$ to their first-order derivative, the transfer function becomes:
\begin{equation}\label{TF}
	\begin{aligned}
		H(k_x,k_y)=\frac{1}{k_0}[-e_{i}^xe_{r}^{x*} (\frac{\partial r_p}{\partial \theta_i}+\frac{k_0}{k_x}r_{p0})+e_{i}^ye_{r}^{y*} (\frac{\partial r_s}{\partial \theta_i}\\+\frac{k_0}{k_x}r_{s0})]k_x
		-\frac{cot(\theta_i)}{k_0}(e_{i}^ye_{r}^{x*}+e_{i}^xe_{r}^{y*})(r_{p0}+r_{s0})k_y,
	\end{aligned}
\end{equation}
where $r_{p0}$ and $r_{s0}$ are the reflection coefficients for p- and s- polarized plane waves for the central wavevector $\boldsymbol{k_0}$, $\frac{\partial r_{p(s)}}{\partial \theta_i}$is the first-order derivative of the reflection coefficient with respect to the incident angle $\theta_i$. 

We take the first type $s-$ polarized topological singularity as an example, based on the zero scattering property, the condition $r_{s0}=0$ is satisfied. Here we set the incident polarization and the reflected polarization as $\boldsymbol{e}{_{i}}=(0,1)^T$ and $\boldsymbol{e}{_{r}}=(0,1)^T$, the transfer function $H$ becomes:  
\begin{equation}\label{TFx}
	\begin{aligned}
		H(k_x,k_y)=\frac{1}{k_0} \frac{\partial r_s}{\partial \theta_i}  k_x.
	\end{aligned}
\end{equation}

The second step to achieve differential imaging involves transforming the $k$-space equations into the real space, e.g.  $k_x$(or $k_y$) is transformed into $\frac{\partial}{\partial x}$(or $\frac{\partial}{\partial y}$). 
Then, the relation in Eq. (\ref{Er}) is transformed into:
 \begin{equation}
 E_{r}(x,y)=\widehat{D}E_{i}(x,y),
 \end{equation} 
in real space, where $\widehat{D}$ is a differential operator. For example, Eq. (\ref{TFx}) corresponds $\widehat{D}=\frac{\partial}{ \partial x}$, which represents 1D first-order optical differential imaging along $x-$ direction\cite{zhu2019generalized}. 

From the derivation of the transfer function $H$, 
we can see that, for oblique incident cases, $H$ can be expanded as the polynomials of $k_i$ terms by choosing polarizations and with certain conditions, where $i=x$, $y$. With those theory in mind, we will show how to realize these transfer function $H$ in real optical systems.   
\begin{figure*}[htb]
	\includegraphics[width=\linewidth]{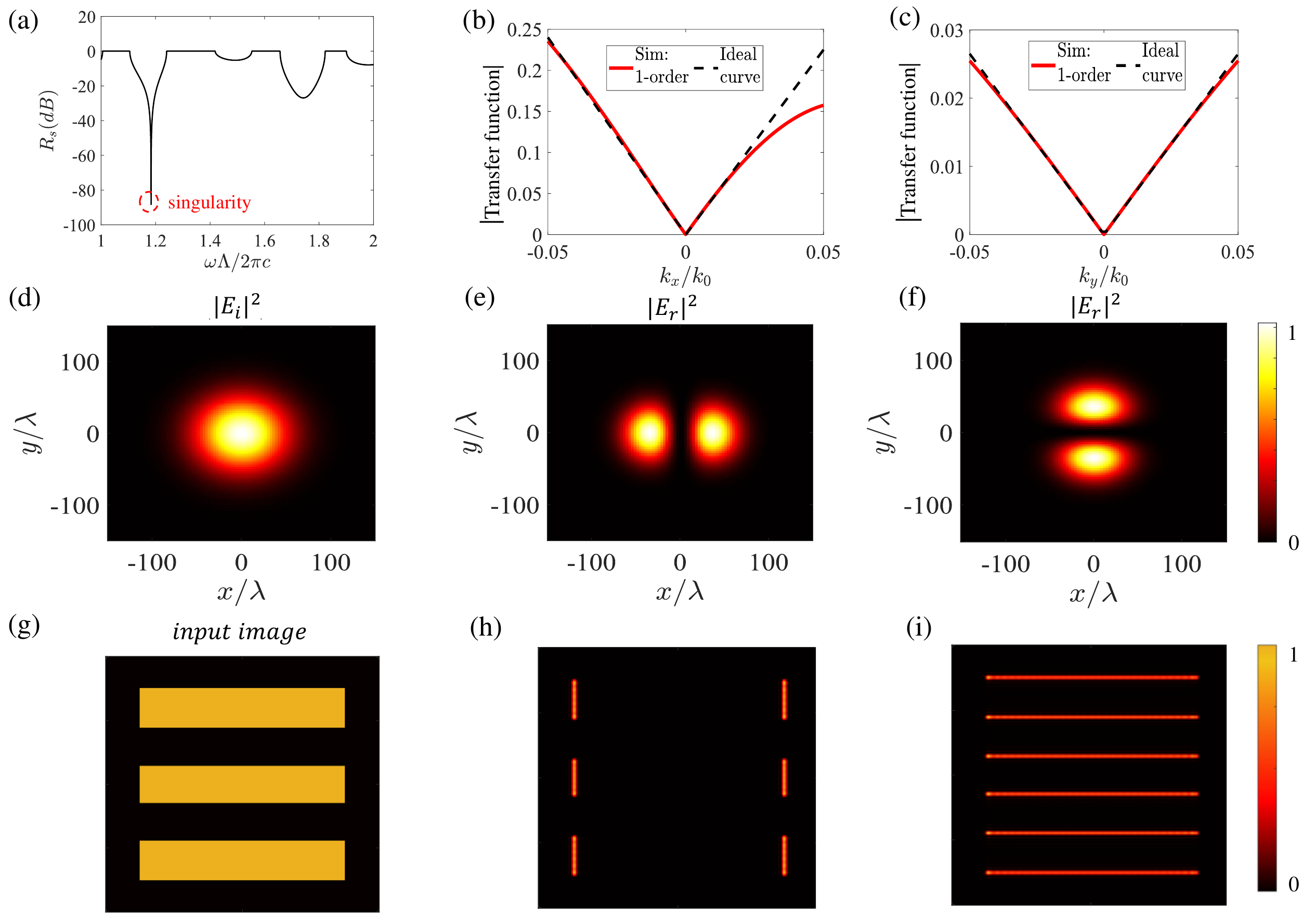}
	\caption{\label{fig:1Ddiff} (a)Reflectivity (dB) of ABCBA-kind PhC. There is a singularity at  the frequency $\frac{\omega \Lambda}{2\pi c}$=1.18838, which is circled by red dashed line. (b) The red solid line represent the transfer function for $k_y=0$. The black dashed line represents a ideal liner function. (c) The transfer function for $k_x=0$.  (d)Real-space intensity of an incident Gaussian beam with a beam waist of $50\lambda$. (e)-(f)Intensity distributions of reflected light fields of the Gaussian beam with the polarizer $\boldsymbol{e}_{r}=(1,0)^T$ and $\boldsymbol{e}_{r}=(0,1)^T$ respectively. (g)The amplitude object for simulation which consist three rectangular objects. (h)-(i) The simulation results of diffenential imaging. The intensity has been normalized.}
\end{figure*}
\subsection{First-order differential imaging with the first type topological singularity}
In this subsection, we will show that with the first type topological singularities in ABCBA-kind PhCs, 1D first-order optical differential imaging along $x$ or $y$ direction can be realized.  

First, we will derive the form of the transfer function $H$ and study the relationship between topological singularities of 1D ABCBA-kind PhCs and the differential imaging effects under oblique incidence. As we discussed in Eq. (\ref{TFx}), based on same system, we also can achieve differential imaging along $y$- direction by simply tuning the direction of polarizer from $\boldsymbol{e}{_{r}}=(0,1)^T$ to $\boldsymbol{e}{_{r}}=(1,0)^T$. Then the transfer function $H$ can be obtained as:
\begin{equation}\label{TFyy}
	\begin{aligned}
		H(k_x,k_y)=-\frac{cot(\theta_i)}{k_0}(r_{p0}+r_{s0})k_y.
	\end{aligned}
\end{equation}

We depict the reflectivity versus the normalized frequency of semi-infinite PhC in Fig. \ref{fig:1Ddiff}(a), here, the incident angle $\theta_i$ is chosen as $30^{\circ}$. Obviously, there is a low reflection point at the frequency $\frac{\omega \Lambda}{2\pi c}$=1.1883, which represents the position of the topological singularity. For real devices, the 1D PhC is generally with a finite total cell number $N$. For this work, we choose $N$ as $10$ to examine the differential imaging effects, since $N=10$ is large enough and the larger $N$ will show the similar effects. 

With all these parameters, the transfer function $H$ of our system can be calculated by transfer matrix method(TMM) numerically and the results are shown in Fig. \ref{fig:1Ddiff}(b)and \ref{fig:1Ddiff}(c). We depict the transfer function which is indicated by red solid line. We further fit the transfer function with an ideal quadratic function as indicated by the black dashed line. The results of the transfer function are almost linear functions for small $k_i$ where $i=x, y$ and agree very well with our theoretical prediction in Eq.~(\ref{TFx}) and Eq.~(\ref{TFyy}).


Next, to examine the differential imaging effects, we numerically calculate the reflected field with different incident beams. Initially, we consider the incident beam is a Gaussian beam. Fig. \ref{fig:1Ddiff}(d) shows the intensity distribution of the input Gaussian beam. The imaging results for the reflected polarization direction as $\boldsymbol{e}{_{r}}=(0,1)^T$ and $\boldsymbol{e}{_{r}}=(1,0)^T$ are shown in Fig. \ref{fig:1Ddiff}(e) and \ref{fig:1Ddiff}(f), respectively. Indeed, the reflected beam exhibits a Hermite-Gaussian one with a minimum amplitude at $x=0$ or $y=0$, which corresponds to the 1D first-order $x$-direction or $y$-direction differential imaging for the incident Gaussian beam\cite{wen2024tunable}.

Subsequently, we examine the second case where the incident beam carries complicated image information. Specifically, as shown in Fig. \ref{fig:1Ddiff}(g), the incident image consists of three rectangular objects. Numerically, we can obtain the differential imaging by following steps. First, we obtain the components of plane waves of the input image in Fig. \ref{fig:1Ddiff}(g) by Fourier transformation. Second, the reflected plane waves are obtained numerically by the TMM. Third, we choose the certain polarization of the reflected plane waves according the direction of the polarizer. At last, we sum up all these components to obtain the real image of the reflected field, which are shown in Fig. \ref{fig:1Ddiff}(h) and \ref{fig:1Ddiff}(i).  From Fig. \ref{fig:1Ddiff}(h) and \ref{fig:1Ddiff}(i), it can be found that the reflected image clearly present the edges of the input image along the $x$-direction and $y$-direction,  respectively, which indicates the first-order differential imaging along $x-$ and $y-$ direction has been achieved.  In conclusion, the strict numerical results are in excellent agreement with the theoretical predictions.
As the topological singularity evolving with different incident angle, we can achieve the first-order differential imaging at almost arbitrary angle for the first time, which can only be realized at Brewster angle with $x$-polarized incidence in previous work. This facilitates the design of differential imaging devices at arbitrary angles.

\subsection{Second-order $\partial^2 /\partial x \partial y$ type differential imaging with the second type topological singularity}
\begin{figure}[h]
	\includegraphics[width=1\linewidth]{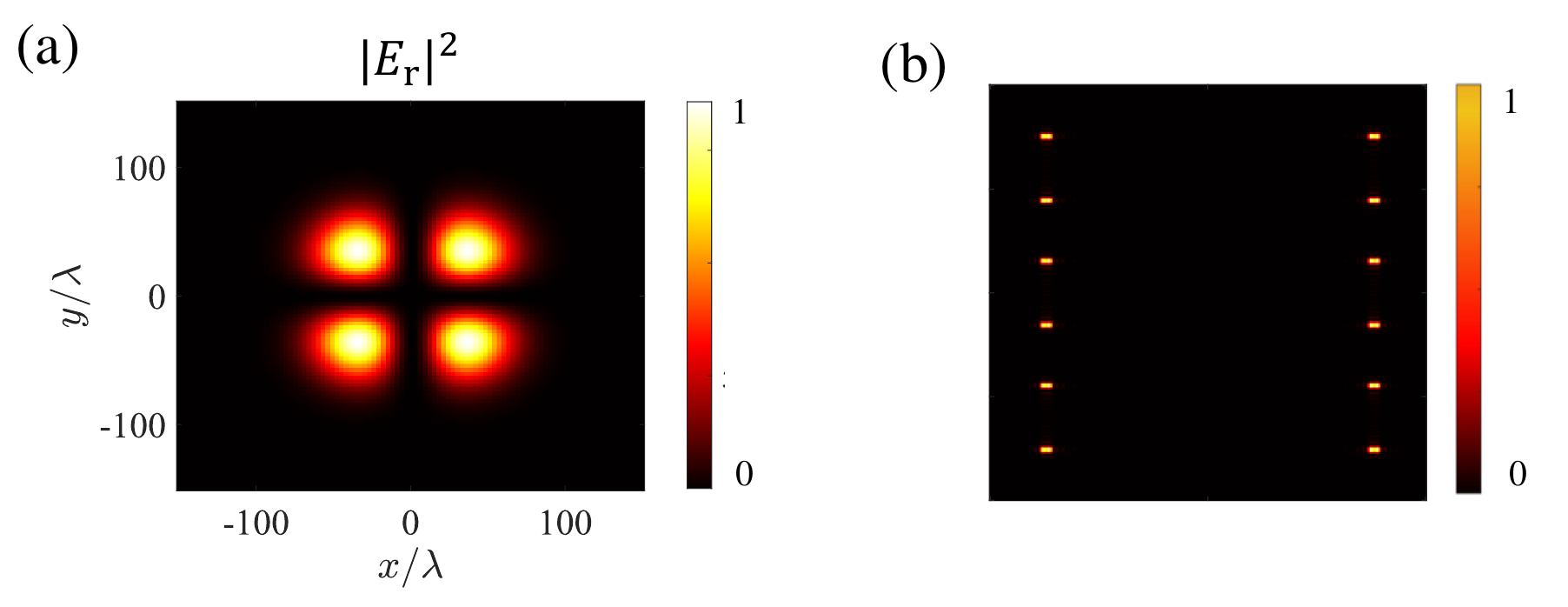}
	\caption{\label{fig:kxky} The numerical calculation results of $\partial^2 /\partial x \partial y$ type differential imaging (a)The intensity distributions of the reflected light fields of the Gaussian beam. (b)The output results of the input image which consist three rectangular objects.}
\end{figure}

Besides the first-order differential imaging realized with the first type topological singularity, we surprisingly find that at the frequencies of the second type topological singularities of ABCBA-kind PhCs, we can achieve a different type differential imaging, i.e. $\partial^2 /\partial x \partial y$ type. 
First, we will derive the form of the transfer function $H$ based on Eq. (\ref{second type}) and Eq. (\ref{TF}).

According to Eq. (\ref{second type}), for the second type topological singularity, the accumulated phase of $p-$polarized and $s-$polarized light in layers $B$ and $C$ are identical. Therefore, the condition of the second type topological singularities can be simultaneously satisfied for $p-$polarized and $s-$polarized light. As a result, the reflection coefficients  for $p-$ and $s-$ polarized light for the central wavevector satisfy:
\begin{equation}\label{rp=rs}
	r_{p0}=r_{s0}=0.
\end{equation}

After substituting  Eq.~(\ref{rp=rs}) into Eq.~(\ref{TF}), at the frequency of the second type topological singularity, the transfer function $H$ becomes:
\begin{equation}\label{TFxy}
	\begin{aligned}
		H(k_x,k_y)=-\frac{cot(\theta_i)}{k_0}(\frac{\partial r_s}{\partial \theta_i}+\frac{\partial r_p}{\partial \theta_i})k_xk_y,
	\end{aligned}
\end{equation}
here, the incident and the reflected polarization are chosen as $\boldsymbol{e}{_{i}}=(0,1)^T$ and $\boldsymbol{e}{_{r}}=(1,0)^T$ respectively. The differential operator corresponding to the transfer function in Eq. (\ref{TFxy}) is:  
\begin{equation}\label{Dkxky}
	\widehat{D}= \frac{\partial^2}{\partial x \partial y},
\end{equation}
which corresponds to $\partial^2 /\partial x \partial y$ type differential imaging effect\cite{shou2022realization}.
We then numerically examine the differential imaging effect. For this purpose, we choose the incident angle as $50^{\circ}$ and the frequency of the first type topological singularity as $\frac{\omega \Lambda}{2\pi c}=5.1$. The results for the differential imaging $\partial^2 /\partial x \partial y$ of a Gaussian beam are shown in Fig.~\ref{fig:kxky}(a). The reflected image exhibit the Hermite-Gaussian modes $HG_{11}$ pattern, which is in agreement with the theoretical prediction\cite{tu2025inverse}. 

Subsequently, we examine the differential imaging using an incident image consisting of three rectangular objects, as depicted in Fig. \ref{fig:1Ddiff} (g). The imaging results are shown in Fig. \ref{fig:kxky}(b). The reflected image exhibits only dots at the corner, which results from first-order differential imaging along $x$ and $y$ direction, respectively, i.e. firstly, the first-order $x-$ direction differential imaging transforms the square into two vertical lines, then $y-$ direction differential imaging converts these lines into individual dots\cite{tu2025inverse}. Since the second type topological singularity for ABCBA-kind PhC can be degenerated into the topological singularity for ABA-kind PhC. This type differential imaging effect can also be realized in ABA-kind PhC, detailed description can be found in Appendix .\ref{ABA}.
We note that, to the best of our knowledge, the $\partial^2 /\partial x \partial y$ type differential imaging is firstly realized in PhCs without the use of cascading. This provides a good platform for achieving this type differential imaging.
\subsection{Deep subwavelength differential imaging with the first type topological singularities under oblique incidence.} \label{deep}
\begin{figure}[htb]
	\includegraphics[width=1\linewidth]{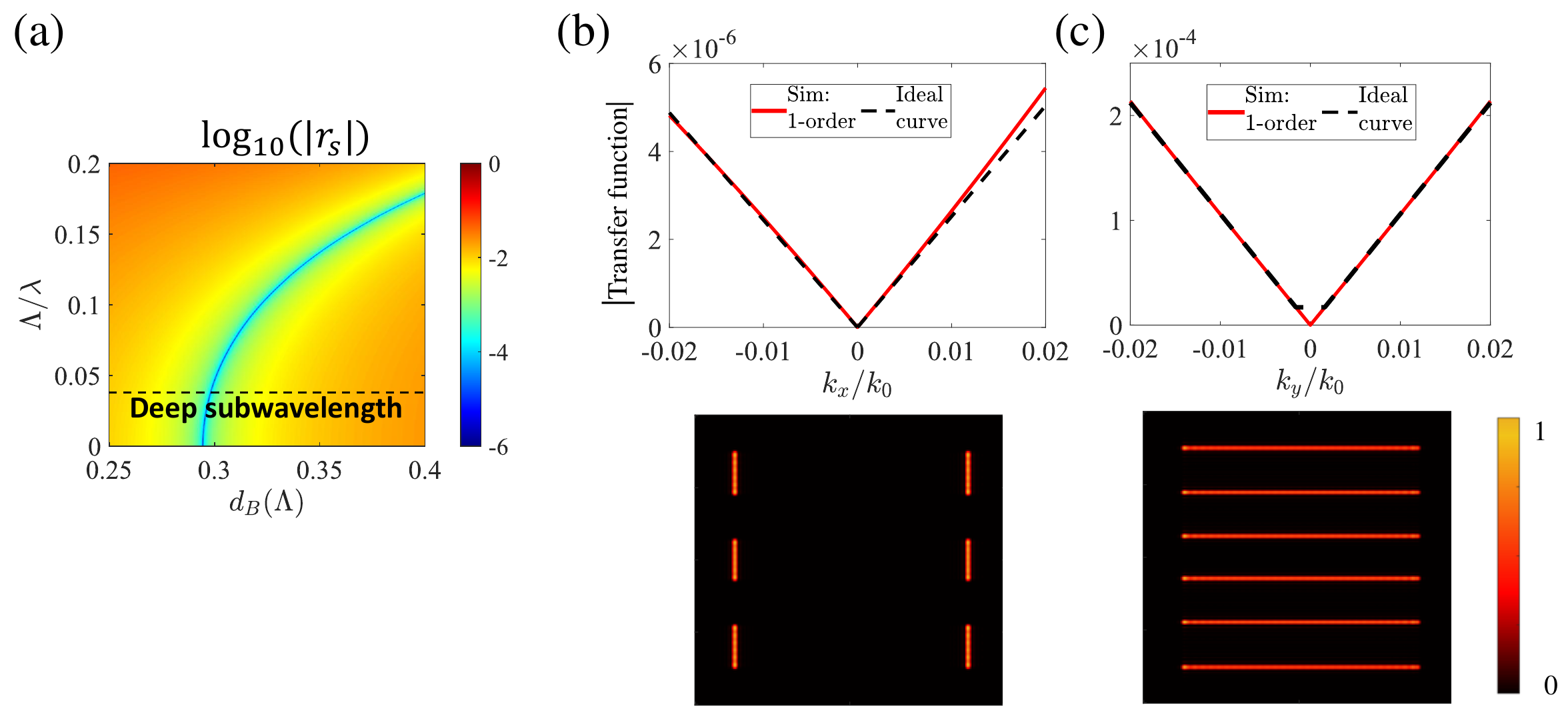}
	\caption{\label{fig:deepsub} (a)The reflection coefficients $log_{10}(|r_s|)$ versus the length of layer-B $d_B$ and the wavelength $\lambda$ of semi-infinite PhC. The deep blue line is the trajectory of the first type topological singularity. The deep subwavelength region is defined as $\lambda>20\Lambda$, as indicated in the region below the black dashed line. (b-c)are the transfer function of ABCBA-kind PhCs at the frequency of the first type topological singularity(upper panel) and the numerical simulated differential imaging results of the first-order differential imaging(lower panel).}
\end{figure}
It is well known that deep subwavelength optical devices, which operate at scales smaller than the wavelength of light they interact with, enable the integration of optical devices into tight spaces, such as smartphones. However, due to the limitations of the effective medium theory, the research on differential imaging at deep subwavelength scales is relatively limited at present. Recently, Xiong et al. \cite{xiong2021resonance} successfully move the first type topological singularity to deep subwavelength region. This interesting phenomenon provides a feasible scheme to realize differential imaging in deep subwavelength region.
In this subsection, we will demonstrate that first-order differential imaging in deep subwavelength can be realize in ABCBA-kind PhCs under oblique incidence by shifting the frequency of the first type topological singularities to the deep subwavelength region.
Next, we will examine the differential imaging effect following the same steps mentioned above. 
To achieve the emergence of the topological singularity in deep subwavelength region, in this model, the relative permittivity of the five layers are set as $\varepsilon_A=2.1$, $\varepsilon_B=3.24$, $\varepsilon_C=1.44$, $\varepsilon_B=3.24$, $\varepsilon_A=2.1$ and the length of the layers are set as $d_A=0.2\Lambda$, $d_C=\Lambda-d_A-d_B$, respectively. In Fig.~\ref{fig:deepsub}(a), we show the reflection coefficients $log_{10}(|r_s|)$ versus the length of layer-B $d_B$ and the wavelength $\lambda$ of semi-infinite PhC. As can be seen in Fig.~\ref{fig:deepsub}(a), the deep blue lines, indicating perfect transmission and zero reflection, are the trajectory of topological singularity in \{$d_B,\lambda$\} space. Obviously, with the decrease of $d_B$, the singularity will move towards lower frequency (longer wavelength) region and it reaches zero frequency at $d_B=0.2944a$, further decreasing $d_B$, the topological singularity will evolve into the pure imaginary frequency region. In this work, the deep subwavelength region is defined as $\lambda>20\Lambda$, as indicated in the region below the black dashed line.
To further examine the differential imaging effect, we set the cell number $N=10$, the incident angle $\theta_i=10^{\circ}$ and the length of the layer-B $d_B=0.295\Lambda$. The wavelength of topological singularity is chosen as $\lambda=38.98\Lambda$.
With all these parameters, the transfer function $H$ of our system can be calculated numerically and the results are shown in the upper panel of Figs. \ref{fig:deepsub}(b) and \ref{fig:deepsub}(c). It shows that the transfer function(red solid line) is well fit with the ideal linear function(black dashed line). 

We then verify the differential imaging with three rectangular shaped objects as the input image. Following the same steps mentioned above, the differential imaging results in the lower panel of Fig. \ref{fig:deepsub}(b) and (c) clearly present the edges of the input image along the $x$-direction and $y$-direction, respectively. Therefore, first-order differential imaging is realized in deep subwavelength region with the first type topological singularity.

\section{HIGHER-ORDER DIFFERENTIAL IMAGING FROM TOPOLOGICAL SINGULARITIES OF PHCS UNDER NORMAL INCIDENCE}
Higher-order differential imaging often yields sharper boundaries compared to lower-order differential imaging, thereby significantly enhancing image edge detection \cite{long2021isotropic}. In this section, we show that when a linearly polarized beam which satisfies both LBWC and NSC is \textbf{normally} incident on the surface of the PhC and the reflected beam, after passing a polarizer, could present radial second-order or even fourth-order differential imaging. The transfer function of this differential imaging takes the form:
\begin{equation}
	H\propto k_\rho^n,
\end{equation}
  where n is a positive integer. Specifically, second-order and even fourth-order differential imaging can be achieved with different types of topological singularities of ABCBA-kind PhCs. This second-order and fourth-order radial differential imaging effects can be theoretically analyzed through the transfer function around the topological singularity. We will show that the transfer function $H$ with the form as $k_\rho^n$ can be easily achieved around topological singularities of 1D PhCs with certain conditions. Hence the higher-order differential imaging can be generally designed based on our mechanism.  
  \begin{figure*}[htb]
	\includegraphics[width=1\linewidth]{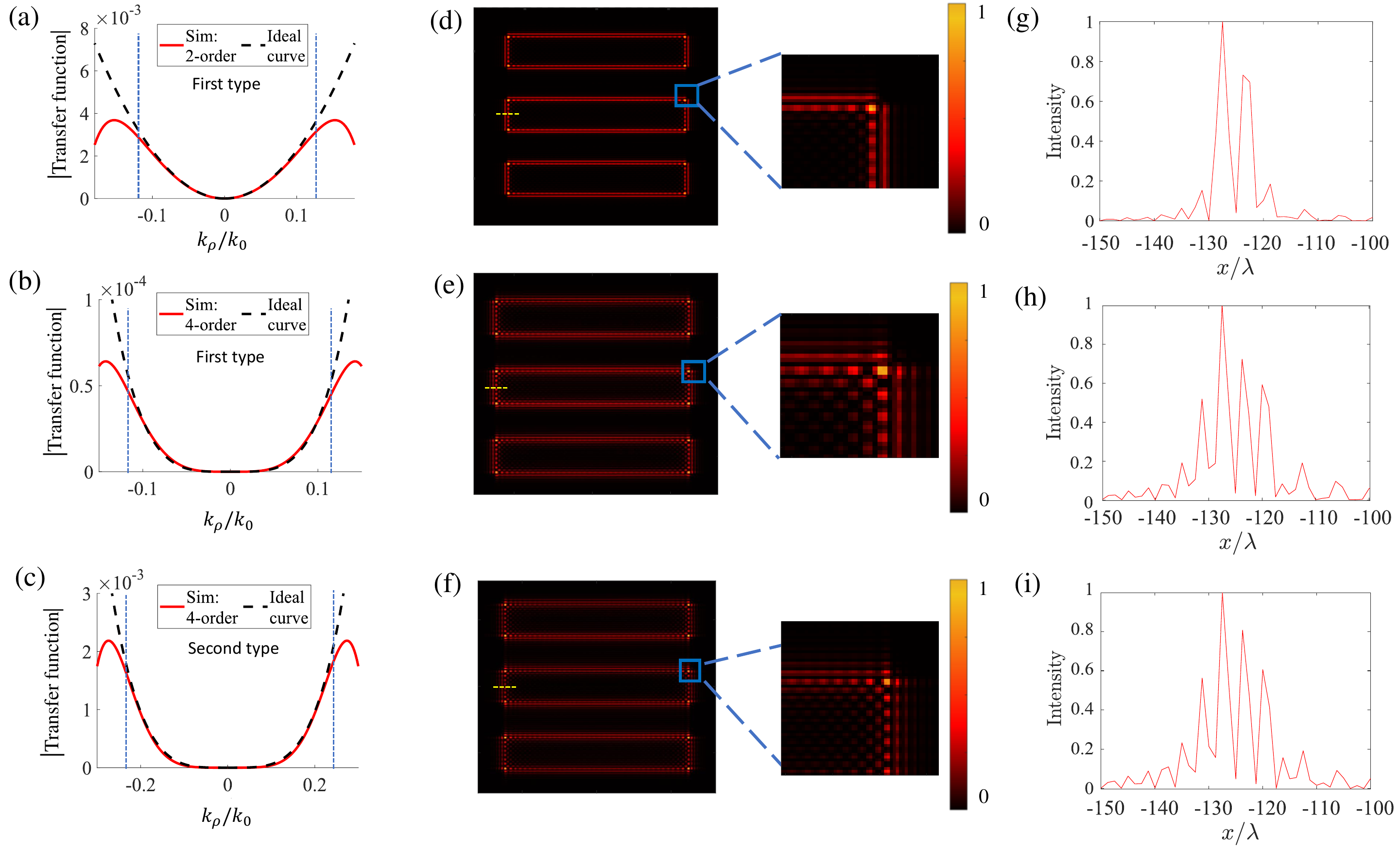}
	\caption{\label{fig:2D4diff} Differential imaging under normal incidence. (a-b)The transfer function magnitudes as functions of radial wave vector $k_\rho$ for the first type topological singularity in ABCBA-kind PhCs(red solid line) and the corresponding ideal curve(black dashed line). (c)The transfer function magnitudes as functions of radial wave vector $k_\rho$ for the second type topological singularity in ABCBA-kind PhCs(red solid line) and the corresponding ideal curve(black dashed line). (d-f)Imaging result of the input image which consist three rectangular objects, the inset shows the detailed image of the right edge. (g-i) Normalized intensity along the yellow dashed line of the image in (d-f).}
\end{figure*}

\subsection{The transfer function of the system under normal incidence}
For the normal incident cases, the transformation matrix $M$ differs from that of the oblique incident cases and can be expressed as:
\begin{equation}\label{eq17}
	\boldsymbol{M}=\left[
	\begin{array}{cc}
		-cos(\varphi)& sin(\varphi)\\
		sin(\varphi)cos(\theta_i)&cos(\varphi)cos(\theta_i)
	\end{array}
	\right],
\end{equation}
where $\theta_i=sin^{-1}(\sqrt{k_x^2+k_y^2}/k_0)$ is the incident angle and $\varphi=tan^{-1}(k_y/k_x)$ is the azimuthal angle\cite{long2021isotropic}. 
After substituting Eq.~(\ref{eq17}) into Eq.~(\ref{H}), the new transfer function $H(\varphi,k_\rho)$ can be derived as :
\begin{equation}\label{eq18}
	\begin{aligned}
		H(\varphi,k_\rho)=\frac{1}{k_0}(e_{i}^xe_{r}^{x*}(r_scos^2(\varphi)+r_psin^2(\varphi))\\
		+(e_{i}^xe_{r}^{y*}+e_{i}^ye_{r}^{x*}) (\frac{sin(2\varphi)}{2}(r_p-r_s))\\
		+(e_{i}^ye_{r}^{y*}(r_pcos^2(\varphi)+r_ssin^2(\varphi)).
	\end{aligned}
\end{equation}
 If we choose $\boldsymbol{e}_{{i}}=[0,1]^T$ and the reflected polarization to be $\boldsymbol{e}_{{r}}=[1,0]^T$, the transfer function is obtained as:
 \begin{equation}\label{eq18k}
 	\begin{aligned}
 		H(\varphi,k_\rho)= \frac{sin(2\varphi)}{2}(r_p-r_s).
 	\end{aligned}
 \end{equation}
 Under the normal incidence condition, considering the in plane rotational symmetry of the PhCs structure, the Taylor expansions of the reflection coefficients $r_p$ and $r_s$ contain only even-order terms of $k_\rho$. So, the difference in reflection coefficient $r_p-r_s$ can be expressed as :
  \begin{equation}\label{rp-rs}
 	\begin{aligned}
 	r_p-r_s=C_1  k_\rho^2+C_2 k_\rho^4+......+C_n k_\rho^{2n} ,
 	\end{aligned}
 \end{equation}
  where $C_n$ are the coefficients of the expansion terms. Generally, in common cases, the term $C_1  k_\rho^2$ is the dominant one and other expansion terms can be neglected. Then the transfer function can be written as:
   \begin{equation}\label{c1}
   	\begin{aligned}
   		H(\varphi,k_\rho)= \frac{sin(2\varphi)}{2}C_1  k_\rho^2.
   	\end{aligned}
   \end{equation}
 Consequently, when the azimuth angle $\varphi$ is fixed as a constant, the transfer function $H$ is a quadratic function related to $k_\rho$, i.e.,
  \begin{equation}\label{c11}
 	\begin{aligned}
 	H\propto k_\rho^2.
 	\end{aligned}
 \end{equation}
 It is an essential condition to realize radial second-order differential imaging. Physically, when we fixed the azimuth angle $\varphi$, it implies that the incident light in Fourier space has only the field components along the direction with the angle $\varphi$ to $k_x$ axis. Interestingly, in some special cases, we will show $C_1=0$ near the topological singularities of PhCs. So, the term $C_2 k_\rho^4$ is dominant. When the azimuth angle $\varphi$ is fixed as a constant, the transfer function $H$ is a quartic function related to $k_\rho$:
  \begin{equation}\label{c2}
  	\begin{aligned}
  		H\propto k_\rho^4.
  	\end{aligned}
  \end{equation}
 From the derivation of the transfer function $H$, 
 we can see that, for normal incident cases, $H$ can be expressed as the polynomials of $k_\rho$ terms by choosing polarizations and with certain conditions. Therefore, the $2n$th-order radial differential imaging where $E_{r}(x,y)\propto \frac{\partial ^{2n}E_{i}(x,y)}{\partial ^{2n}\rho}$($\rho=\sqrt{(x^2+y^2)}$) in real space, can be realized. With those theory in mind, we will show how to realize these transfer function $H$ in real optical systems.   
  
   

\subsection{2D second-order and fourth-order radial differential imaging with the first type singularities.}

In this subsection, we demonstrate that at the frequency of the first type singularities in ABCBA-kind PhCs, radial second-order differential imaging can be generally realized under normal incidence without fine tuning of the structural parameters. Furthermore, it is possible to achieve radial fourth-order differential imaging by tuning the structural or material parameters. 

Firstly, we will derive the form of the transfer function $H$ theoretically for N unit cells ABCBA-kind PhCs at the frequency of first type singularities. 
For ABCBA-kind PhCs,  the transfer matrix $T$ of a single unit cell PhC can be written as:
\begin{equation}\label{eqT}
T=\left[
\begin{array}{cc}
	t_{11} & t_{12} \\
	t_{12}^* & t_{11}^*
\end{array}
\right].
\end{equation}
Eq. \ref{eqT} can be reformulated in the Pauli matrix form as:
\begin{equation}
	T = \frac{t_{11}+t_{11}^*}{2} \hat{\sigma}_0 + \frac{t_{12}+t_{12}^*}{2}\hat{\sigma}_x + \frac{t_{12}-t_{12}^*}{2i} \hat{\sigma}_y + \frac{t_{11}-t_{11}^*}{2} \hat{\sigma}_z,
\end{equation}
where $\hat{\sigma}_i$ (i = $x$, $y$, $z$) is the Pauli matrix, $\hat{\sigma}_0$ is a $2\times2$ identity matrix. At the frequency of topological singularity, we have $t_{12}=0$\cite{li2019singularity}. Under the normal incidence condition, considering the in plane rotational symmetry of the PhCs structure, in $k-$ space, each element of our matrix is even-symmetric about $k_\rho=0$. Therefore, the Taylor expansions of each element of the transfer matrix $T$ around $k_\rho=0$ contains only even order terms of $k_\rho$. Consequently, we can expand each element of the transfer matrix $T$ around $k_\rho=0$ as:
\begin{equation}\label{TM}
T = a \hat{\sigma}_0 + (b k_\rho^2+ck_\rho^4)\hat{\sigma}_x + (d k_\rho^2+ek_\rho^4) \hat{\sigma}_y + f \hat{\sigma}_z,
\end{equation}
$a,b,c,d,e,f$ are the coefficients of the expanding terms. According to Eq.(\ref{eqT}) and Eq.(\ref{TM}), for a single unit cell PhC, the reflection coefficients can be expressed as: 
\begin{equation}\label{r}
r_{s(p)}=t_{12}/t_{11}=[(b-id)_{s(p)}k_\rho^2+(c-ie)_{s(p)}k_\rho^4]/(a+f),
\end{equation}
where $s$ and $p$ represent $s-$ and $p-$ polarized wave.
Furthermore, for N unit cells PhC the transfer matrix $T$ can be written as:
\begin{equation}
	T=\left[
	\begin{array}{cc}
		t_{11} & t_{12} \\
		t_{12}^* & t_{11}^*
	\end{array}
	\right]^N.
\end{equation}
Then the reflection coefficients can be easily obtained as:
\begin{equation}\label{rn}
r_{s(p)}=\beta[(b-id)_{s(p)}k_\rho^2+(c-ie)_{s(p)}k_\rho^4]/(a+f)^N,
\end{equation}
here $\beta$ is a constant. According to Eq.~(\ref{eq18k}) and Eq.~(\ref{rn}), we can get the form of the transfer function $H$ for N unit cells ABCBA-kind PhCs as:
\begin{equation}\label{Hps}
\begin{split}
	H(\varphi,k_\rho) = & \frac{\beta \sin(2\varphi)}{2(a+f)^N} \left[ \left( (b-id)_p - (b-id)_s \right) k_\rho^2 \right. \\
	& \left. + \left( (c-ie)_p - (c-ie)_s \right) k_\rho^4 \right].
\end{split}
\end{equation}
Here we set $\boldsymbol{e}_{{i}}=[0,1]^T$ and the reflected polarization $\boldsymbol{e}_{{r}}=[1,0]^T$.
In Appendix \ref{ABCBA}, we will show in ABCBA-kind PhCs at the frequency of the first type topological singularity, $(b-id)_p \neq (b-id)_s$, then, the transfer function is given by:
\begin{equation}
H\propto sin(2\varphi)k_\rho^2.
\end{equation}  
Consequently, when the azimuth angle $\varphi$ is fixed as a constant, radial second-order differential imaging can be realized. Next, we will verify whether the transfer function is a quadratic function of the radial wave vector around the first type topological singularity. Without loss of generality, we choose a PhC with 10 unit cells as our design(more unit cell follow the same rule). We numerically calculate the transfer function at the frequency of the first type topological singularity($\frac{\omega \Lambda}{2\pi c}=1.745$). Here, the azimuth angle is fixed as $\varphi=45^{\circ}$. In Fig. \ref{fig:2D4diff}(a), we depict the transfer function which is indicated by red solid line. We further fit the transfer function with an ideal quadratic function as indicated by the black dashed line. The agreement between the transfer function and the ideal quadratic function within the radial wave vector up to $0.13k_0$ confirms the second-order proportional relationship of the transfer function along the radial direction. 

Furthermore, we use numerical experiments to verify that our system indeed possess second-order differential imaging. The incident beam is the same as Fig. \ref{fig:1Ddiff}(g), which consists of three rectangular shaped objects. Numerically, we can obtain the differential imaging results by following steps. First, we find the components of plane waves of the input image by Fourier transformation. Second, we transform the plane wave components from the Cartesian coordinate($k_x,k_y$) system to the polar coordinate system($k_\rho,\varphi$). Third, we use TMM to numerically calculate the reflected plane waves. At last, we transform the coordinate back to Cartesian one and sum up all the component to obtain the real image of the reflected field. The imaging results are shown in Fig. \ref{fig:2D4diff}(d), the inset shows the detail of the edge. As expected, the results clearly present the edges of the three rectangles. Fig. \ref{fig:2D4diff}(g) shows the normalized intensity profile along the yellow dashed line of the image in Fig. \ref{fig:2D4diff}(d), which exhibits two sharp peaks that are consistent with the theory of second-order differential imaging\cite{huo2024broadband}.

Beyond the second-order differential imaging, according to Eq.~(\ref{Hps}), when we tune the structural or materials parameter to make $(b-id)_p = (b-id)_s$, the fourth-order differential imaging can be realized at the frequency of first type topological singularity. As we tune the refractive index of the second layer to $n_b=1.615$ while keeping other parameters unchanged, the frequency of the first type singularity is given by $\frac{\omega \Lambda}{2\pi c}=1.01$. At this frequency, $(b-id)_p = (b-id)_s$, the transfer function can be obtained as:
\begin{equation}
H\propto sin(2\varphi)k_\rho^4. 
\end{equation}
When the azimuth angle is fixed as a constant, the transfer function $H$ is proportional to $k_\rho^4$, i.e., $H\propto k_\rho^4$. The detailed derivation is presented in Appendix .\ref{ABCBA}. Fourth-order differential imaging can be realized. Following the same steps mentioned above, the transfer function and the field distribution of the reflected image are shown in Fig. \ref{fig:2D4diff}(b) and Fig. \ref{fig:2D4diff}(e), respectively. 
Fig. \ref{fig:2D4diff}(h) shows the normalized intensity profile of the left edge, which exhibits four sharp peaks that are consistent with the theory of fourth-order differential imaging\cite{huo2024broadband}.

In summary, in this subsection we find that radial second-order differential imaging can be generally achieved around the first type topological singularity, and by adjusting the structural or material parameters, radial fourth-order differential imaging can be further realized.

\subsection{2D fourth-order radial differential imaging with the second type singularities.}

In this subsection, we will demonstrate that at the frequency of the second type topological singularities in ABCBA model, radial fourth-order differential imaging effect can be generally achieved which is apparently different to the first type topological singularity(generally second-order). 

The condition of the second type topological singularity under normal incident is $sin(k_{Bz}d_B)=sin(k_{Cz}d_C)=0$. Combining this condition with Eq.~(\ref{Hps}), we have $(b-id)_p=(b-id)_s$. The detailed derivation is presented in Appendix .\ref{ABCBA}. Consequently,  the transfer function can be written as:
\begin{equation}
H\propto sin(2\varphi)k_\rho^4.
\end{equation}   
So, when the azimuth angle $\varphi$ is fixed as a constant, the transfer function $H$ is a quartic function related to $k_\rho$, i.e., $H(\varphi,k_\rho)\propto k_\rho^4$. It is a essential condition to realize radial fourth-order differential imaging. 
Following the same steps mentioned above, we first numerically calculate the transfer function. The frequency of the second type of singularity is given by $\frac{\omega \Lambda}{2\pi c}=1.111$. The result is shown in Fig. \ref{fig:2D4diff}(c), the black dashed line indicate ideal quartic function.  The agreement between the transfer function and the ideal quartic function within the radial wave vector up to $0.24k_0$ confirms the fourth-order proportional relationship of the transfer function along the radial direction. We also numerically calculate the reflected field, the incident beam is the same as Fig. \ref{fig:1Ddiff}(g). The imaging result is shown in Fig. \ref{fig:2D4diff}(f), the inset shows the detail of the edge. As we have expected, the reflected light clearly present the edges of the three rectangles. Fig.~\ref{fig:2D4diff}(i) shows the normalized intensity profile of the left edge, where four closely spaced peaks can be observed. These confirm the realization of fourth-order differential imaging.

In Appendix \ref{ABA}, we show that the fourth-order differential imaging mentioned in this subsection can also be realized in ABA-kind PhC.

\subsection{Deep subwavelength differential imaging with the first type topological singularities under normal incidence.}
\begin{figure}[h]
	\includegraphics[width=1\linewidth]{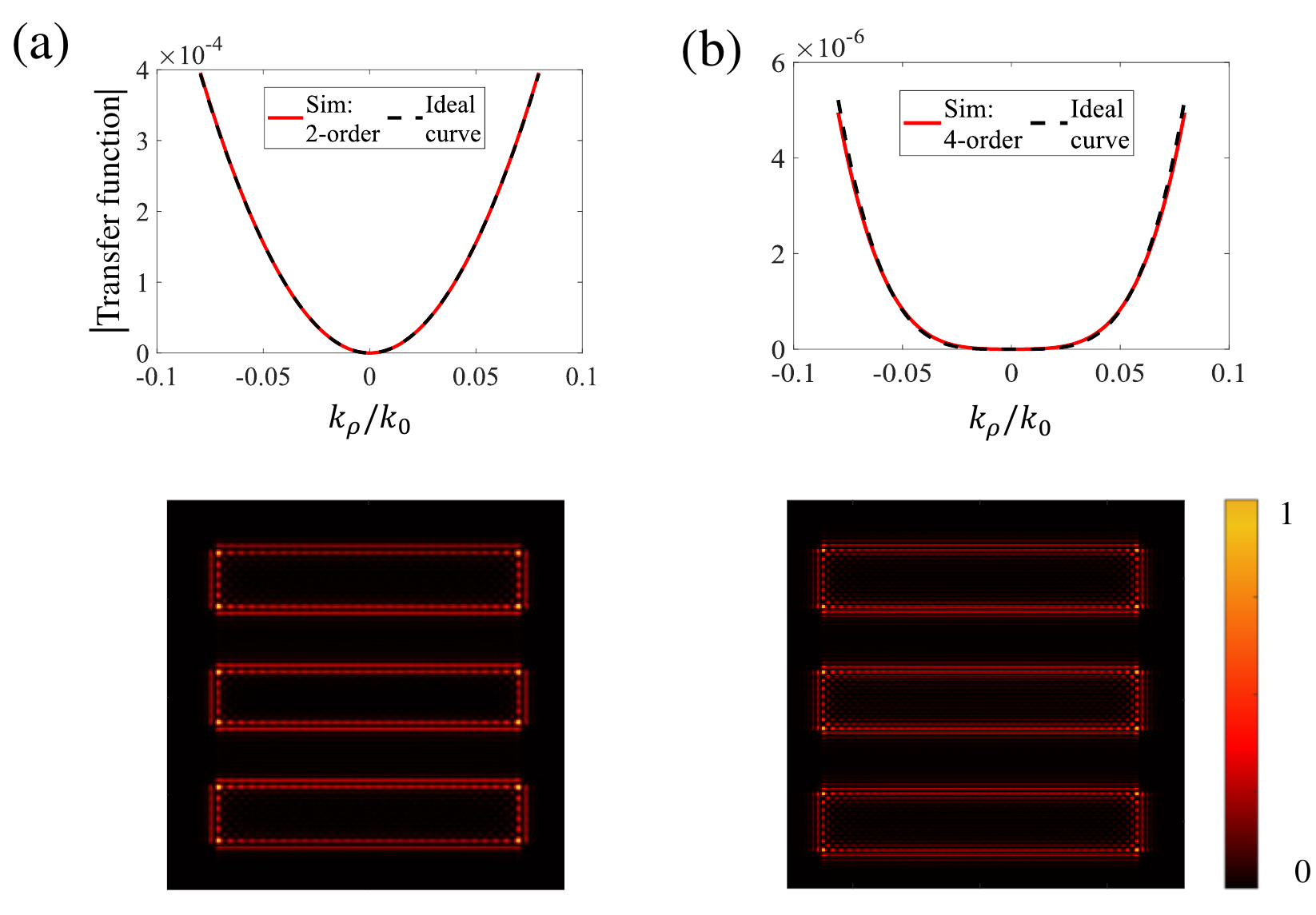}
	\caption{\label{fig:deepsubwavelength} (a)The transfer function of ABCBA-kind PhCs at the frequency of the first type topological singularity under normal incidence and the simulated differential imaging results of the second-order differential imaging(lower panel). The wavelength of the topological singularity we used here is $\lambda=38.98\Lambda$. (b)  The transfer function and the simulated differential imaging results of the fourth-order differential imaging(lower panel). After tuning the layer-B thickness $d_B$ to $0.2985\Lambda$, the wavelength of the topological singularity we used is $\lambda=23.24\Lambda$. }
\end{figure}
In this subsection, we will show second-order and fourth-order radial differential imaging with the first type topological singularities can also be realized in deep subwavelength region. We examine the differential imaging effect following the same steps mentioned above. The model parameter is the same as subsection. \ref{deep}. The result is shown in Fig.~\ref{fig:deepsubwavelength}(a), the wavelength of the  first type topological singularity is chosen as $\lambda=38.99\Lambda$. Second-order differential imaging can be realized.  Also, by tuning the thickness of layer-B $d_B$ to $0.2985\Lambda$ and choosing the wavelength of the first type topological singularity as $\lambda=23.24\Lambda$, we can further realize fourth-order radial differential imaging in deep sub-wavelength, which is proved in Fig.~\ref{fig:deepsubwavelength}(b).
 
 \section{CONCLUSION}
 In summary, we have designed an 1D PhC to achieve multiple-order differential imaging by utilizing the topological of $s-$ or $p-$polarization, that is special perfect transmission modes. We show that for oblique incident cases, with the first type topological singularity, first-order differential imaging in $x-$ and $y-$ direction can be realized. With the second type topological singularity,  $\partial^2 /\partial x \partial y$- type differential imaging can be achieved. For normal incident cases, with the first type topological singularity, radial second-order differential imaging can be realized. Furthermore, when we tune the structural or material parameter, radial fourth-order differential imaging can be realized. We also find that with the second type topological singularity, radial fourth-order differential imaging can be realized without parameter tuning. Additionally, by shifting the first type topological singularity to the deep subwavelength region, the corresponding differential imaging effect can also be realized in this frequency region which fundamentally breaks the limitations of the effective medium theory. Based on these theoretical findings, multi-order, multi-angles, and multi-frequencies differential imaging devices can be theoretically designed. Within the same system,  different order differential imaging can be switched by adjusting the incident frequencies and angles to match different type singularities. Since the frequency of the topological singularity is closely related to the structural parameters, adjusting these parameters can shift the topological singularity to different frequencies, thereby achieving differential imaging effects over a wide frequency range. This work paves the way for exploring the topology of PhCs to control the differential imaging effect and offer key insights for the design of high-order differential imaging devices.

\begin{acknowledgments}
This work is supported by National Natural Science Foundation of China (Grant No. 12174073).
\end{acknowledgments}

\appendix


\begin{figure*}[htb]
	\includegraphics[width=1\linewidth]{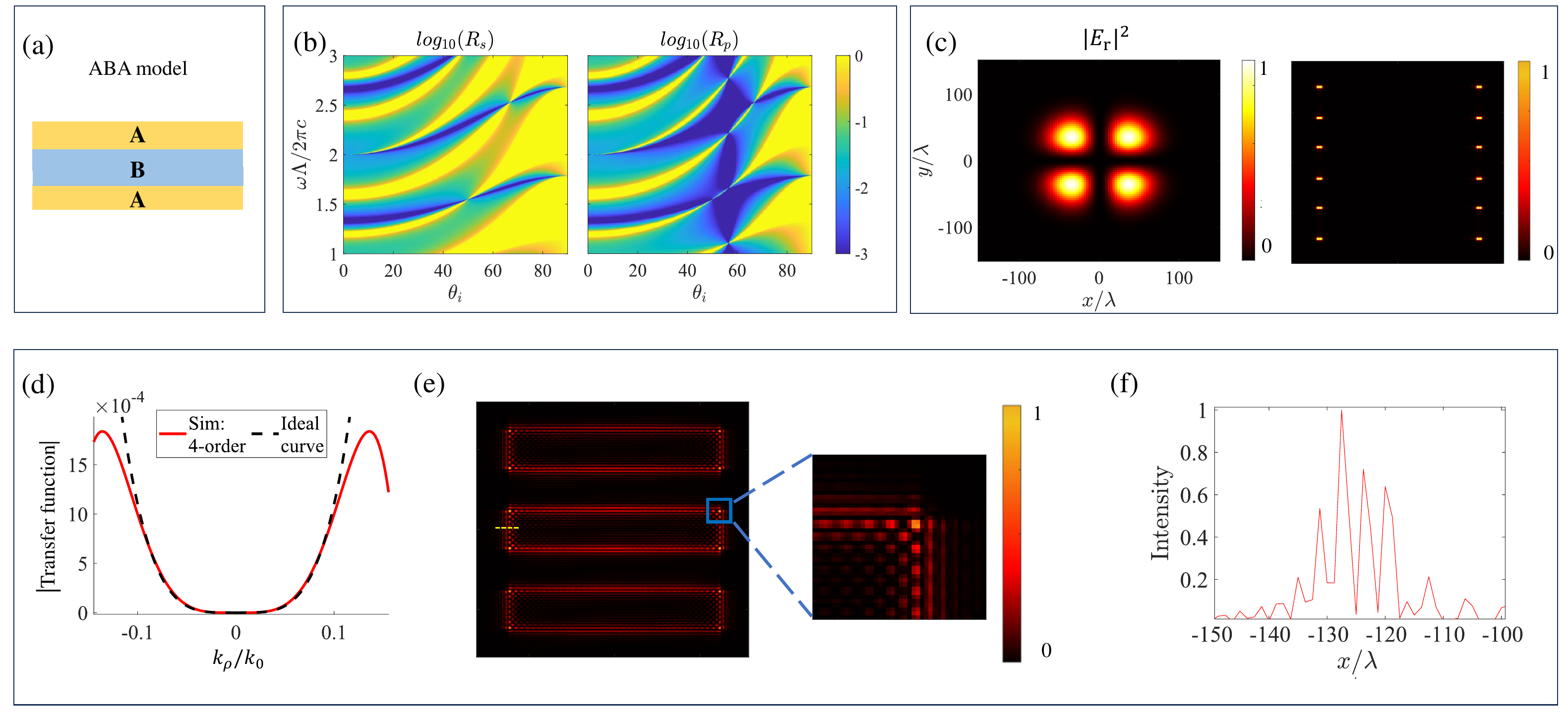}
	\caption{\label{fig16} (a)The unit cell of ABA model. (b) The reflectivity in dB of s-polarization(left panel) and p-polarization(right panel) in $\{\theta_i,\omega\}$space for semi-infinite ABA-kind PhCs. (c) The simulation result of $\partial^2 /\partial x \partial y$ type differential imaging of the Gaussian beam and the input image which consist three rectangular objects. (d) The transfer function of ABA-kind PhCs(red solid line) and the corresponding ideal curve(black dashed line). (e) The simulation result of radial fourth-order differential imaging of the input image which consist three rectangular objects,  the inset shows the detailed image of the right edge. (f) Normalized intensity along the yellow dashed line of the image in (e).}
\end{figure*}

\section{zero scattering properties  at the topological singularities of 1D PhCs}\label{zero}
In this Appendix, we will give a detailed description of the zero scattering property of topological singularity in 1D PhCs. Based on the zero scattering property, we will give the condition of the two types topological singularities in ABCBA-kind PhC.

Firstly, we investigate the of the topological singularities of PhCs. The electric field distribution within this structure can be formulated as follows:
\begin{equation}\label{eq1}
	E(z)=A^+(k_x,k_y)e^{ikz}+A^-(k_x,k_y)e^{-ikz},
\end{equation}
where $A^+$ and $A^-$ are coefficients of forward wave and backward wave, respectively. According to Bloch theory and transfer matrix method \cite{yariv1983optical},\cite{markos2008wave}:
\begin{equation}\label{eq2}
	e^{iK\Lambda}\left[
	\begin{array}{c}
		A^+ \\
		A^-
	\end{array}
	\right]
	=T\left[
	\begin{array}{c}
		A^+ \\
		A^-
	\end{array}
	\right],
\end{equation}
\begin{equation}\label{eq3}
	T=\left[
	\begin{array}{cc}
		t_{11} & t_{12} \\
		t_{21} & t_{22}
	\end{array}
	\right],
\end{equation}
where T is the transfer matrix of a unit cell, K is the Bloch wave vector and $\Lambda$ is the length of a unit cell. Around the topological singularity $A^+\propto\delta K$, $A^-\propto({\delta K})^2$, where $\delta K=(K-K_s)\rightarrow 0$, $K_s$ is the Bloch wave vector of topological singularity\cite{li2019two}, which means that the reflection coefficient:
\begin{equation}\label{|r|}
	 |r|=|\frac{A^-}{A^+}|=0.
\end{equation}
 Therefore the topological singularity can be easily positioned as the zero-scattering point of 1D PhCs.

According to Eq. (\ref{|r|}), in ABCBA-kind PhCs, the topological singularity satisfy the following equation\cite{li2019singularity}:
\begin{equation}\label{sgcs}
	\begin{aligned}
		&F_1 \sin(k_{Cz}d_C) \frac{\cos^2(k_{Bz}d_B)}{2} - F_2 \sin(k_{Cz}d_C) \frac{\sin^2(k_{Bz}d_B)}{2} \\
		&+ F_3 \cos(k_{Cz}d_C) \sin(k_{Bz}d_B)=0,
	\end{aligned}
\end{equation}
where $F_1=\frac{k_{Cz}}{k_{Az}}-\frac{k_{Az}}{k_{Cz}}$, $F_2=\frac{{k_{Bz}}^2}{k_{Az}k_{Cz}}-\frac{k_{Az}k_{Cz}}{{k_{Bz}}^2}$, $F_3=\frac{k_{Bz}}{k_{Az}}-\frac{k_{Az}}{k_{Bz}}$, $F_4=\frac{k_{Cz}}{k_{Bz}}-\frac{k_{Bz}}{k_{Cz}}$ for $s$-wave and $F_1 = \frac{\varepsilon_C k_{Az}}{\varepsilon_A k_{Cz}} - \frac{\varepsilon_A k_{Cz}}{\varepsilon_C k_{Az}}$, $F_2=\frac{{\varepsilon_B}^2k_{Az}k_{Cz}}{\varepsilon_A\varepsilon_C{k_{Bz}}^2}-\frac{\varepsilon_A\varepsilon_C{k_{Bz}}^2}{{\varepsilon_B}^2k_{Az}k_{Cz}}$, $F_3=\frac{\varepsilon_Bk_{Az}}{\varepsilon_Ak_{Bz}}-\frac{\varepsilon_Ak_{Bz}}{\varepsilon_Bk_{Az}}$, $k_{Az}=\sqrt{k_0^2n_A^2-k_\rho^2}$, $k_{Bz}=\sqrt{k_0^2n_B^2-k_\rho^2}$ and $k_{Cz}=\sqrt{k_0^2n_C^2-k_\rho^2}$ are the wave vectors in the z-direction within layer A, B and C, respectively.

According to Eq. (\ref{sgcs}), there are two kinds of solutions, corresponding to two types of topological singularities. The first type satisfies the condition where the left-hand side of the Eq. (\ref{sgc}) equals zero and $sin(k_{Bz}d_B)\neq0$,  $sin(k_{Cz}d_C)\neq0$. The second type satisfies:
\begin{equation}\label{second types}
	sin(k_{Bz}d_B)=sin(k_{Cz}d_C)=0.
\end{equation}


\section{Differential imaging in ABA-kind PhCs}\label{ABA}
Since the second type topological singularity in ABCBA-kind PhCs can be degenerated into the topological singularity in ABA-kind PhCs, in this Appendix, we show that this degeneration allows for the realization of both the $\partial^2 /\partial x \partial y$ type differential imaging and radial fourth-order differential imaging. The cell of 1D ABA-kind PhC is depicted in Fig .(\ref{fig16}(a), which consists three layers in the unit cell with refractive index as $n_A$, $n_B$, $n_A$ and the thickness of three layers are set as $d_A/2$, $d_B$, $d_A/2$, respectively. Without loss of generality, refractive index of layer-$A$ and layer-$B$ are set as $n_A=1$ and $n_B=1.5$, respectively. The thickness of layer-$A$ and layer-$B$ are set as $d_A=d_B=0.5\Lambda$.
Firstly, for oblique incident cases, in ABA-type PhCs, the condition of topological singularities is given by  $sin(k_{Bz}d_B)=0$. Combining with Eq.~(\ref{TF}), if we suppose the reflected polarization is chosen as  $\boldsymbol{e}{_{r}}=(1,0)^T$, at the frequency of topological singularity, the transfer function $H$ becomes:
\begin{equation}\label{TFxys}
	\begin{aligned}
		H(k_x,k_y)=-\frac{cot(\theta_i)}{k_0}(\frac{\partial r_s}{\partial \theta_i}+\frac{\partial r_p}{\partial \theta_i})k_xk_y.
	\end{aligned}
\end{equation}
which corresponds to the $\partial^2 /\partial x \partial y$ type differential imaging.
The results are shown in Fig .\ref{fig16}(c).

In the case of normal incident, the results are shown in Fig .\ref{fig16}(d)-(f), which verify that the radial fourth-order differential imaging can be realized in ABA-kind PhCs. 

\section{Transfer function of N unit cell ABCBA-kind PhCs at the frequency of topological singularity under normal incidence}\label{ABCBA}
In this Appendix, we would like to give the specific form of $(b-id)_s$ and $(b-id)_p$ with different type topological singularity in ABCBA-kind PhC. Furthermore, we will demonstrate that with the first type topological singularity $(b-id)_s\neq(b-id)_p$, and with the second type topological singularity  $(b-id)_s=(b-id)_p$.
 As we shown in Eq. \ref{Hps} in the main text, the transfer function of N unit cells ABCBA-kind PhCs is given by:
\begin{equation}\label{Hpss}
	\begin{split}
		H(\varphi,k_\rho) = & \frac{\beta \sin(2\varphi)}{2(a+f)^N} \left[ \left( (b-id)_p - (b-id)_s \right) k_\rho^2 \right. \\
		& \left. + \left( (c-ie)_p - (c-ie)_s \right) k_\rho^4 \right].
	\end{split}
\end{equation}

Next, we will demonstrate that at the frequency of the first type topological singularity, $(b-id)_s\neq(b-id)_p$. Therefore the transfer function $H$ is proportional to $k_\rho^2$. At the frequency of the second type topological singularity, $(b-id)_s=(b-id)_p$. Therefore the transfer function $H$ is proportional to $k_\rho^4$.
we start from the form of $b-id$ for $s-$ and $p-$ wave respectively. Under the condition of the first type topological singularity, we have:
\begin{equation}\label{b-id}
	\begin{aligned}
		(b-id)_s &= e^{ik_{Az}d_A} \left. \frac{\partial^2 E_s}{\partial k_\rho^2} \right|_{k_\rho=0} \\
		(b-id)_p &= e^{ik_{Az}d_A} \left. \frac{\partial^2 E_p}{\partial k_\rho^2} \right|_{k_\rho=0}
	\end{aligned}
\end{equation}
\begin{equation}\label{Es}
	\begin{split}
		E_s = & \biggl( F_{1s} \sin(k_{Cz}d_C) \cos^2\left(\frac{k_{Bz}d_B}{2}\right) \\
		& - F_{2s} \sin(k_{Cz}d_C) \sin^2\left(\frac{k_{Bz}d_B}{2}\right) \\
		& + F_{3s} \cos(k_{Cz}d_C) \sin(k_{Bz}d_B) \biggr)
	\end{split}
\end{equation}
where  $F_{1s}=\frac{k_{Cz}}{k_{Az}}-\frac{k_{Az}}{k_{Cz}}$, $F_{2s}=\frac{{k_{Bz}}^2}{k_{Az}k_{Cz}}-\frac{k_{Az}k_{Cz}}{{k_{Bz}}^2}$, $F_{3s}=\frac{k_{Bz}}{k_{Az}}-\frac{k_{Az}}{k_{Bz}}$, $F_{4s}=\frac{k_{Cz}}{k_{Bz}}-\frac{k_{Bz}}{k_{Cz}}$.
\begin{equation}\label{Ep}
	\begin{split}
		E_p = & \biggl( F_{1p} \sin(k_{Cz}d_C) \cos^2\left(\frac{k_{Bz}d_B}{2}\right) \\
		& - F_{2p} \sin(k_{Cz}d_C) \sin^2\left(\frac{k_{Bz}d_B}{2}\right) \\
		& + F_{3p} \cos(k_{Cz}d_C) \sin(k_{Bz}d_B) \biggr)
	\end{split}
\end{equation}
where $F_{1p} = \frac{\varepsilon_C k_{Az}}{\varepsilon_A k_{Cz}} - \frac{\varepsilon_A k_{Cz}}{\varepsilon_C k_{Az}}$, $F_{2p}=\frac{{\varepsilon_B}^2k_{Az}k_{Cz}}{\varepsilon_A\varepsilon_C{k_{Bz}}^2}-\frac{\varepsilon_A\varepsilon_C{k_{Bz}}^2}{{\varepsilon_B}^2k_{Az}k_{Cz}}$, $F_{3p}=\frac{\varepsilon_Bk_{Az}}{\varepsilon_Ak_{Bz}}-\frac{\varepsilon_Ak_{Bz}}{\varepsilon_Bk_{Az}}$, $F_{4p}=\frac{\varepsilon_Bk_{Cz}}{\varepsilon_Ck_{Bz}}-\frac{\varepsilon_Ck_{Bz}}{\varepsilon_Bk_{Cz}}$.

According to Eq.~(\ref{sgc}), under the condition of the first type topological singularity, we can prove:
\begin{equation}
\frac{\partial^2 E_s}{\partial k_\rho^2}|_{k_\rho=0} \neq\frac{\partial^2 E_p}{\partial k_\rho^2} |_{k_\rho=0}
\end{equation}
and
\begin{equation}
	(b-id)_s\neq(b-id)_p
\end{equation}
therefore, according to Eq. (\ref{Hps}), the transfer function becomes:
 \begin{equation}
	H\propto sin(2\varphi)k_\rho^2.
\end{equation}  
For the second type topological singularity ($sin(k_{Bz}d_B)=sin(k_{Cz}d_C)=0$), we have:
\begin{figure*}
\begin{equation*}\label{second b-ics}
	\begin{split}
		(b - id)_s = \frac{e^{ik_{Az}d_A}}{2k_0^2} \bigg( \frac{ F_{1s}d_C}{n_C^2} \cos^2\left(\frac{k_{Bz}d_B}{2}\right) 
		\quad - \frac{F_{2s}d_C }{n_C^2} \sin^2\left(\frac{k_{Bz}d_B}{2}\right) + \frac{ F_{3s}d_B }{n_B^2} \cos(k_{Cz}d_C) \bigg)\bigg|_{k_\rho=0}
	\end{split}
\end{equation*}
\end{figure*}

\begin{figure*}
\begin{equation}\label{second b-icp}
	\begin{split}
		(b - id)_p = \frac{ e^{ik_{Az}d_A}}{2k_0^2} \bigg( \frac{ F_{1p}d_C}{n_C^2} \cos^2\left(\frac{k_{Bz}d_B}{2}\right) 
		\quad - \frac{F_{2p}d_C }{n_C^2} \sin^2\left(\frac{k_{Bz}d_B}{2}\right) + \frac{ F_{3p}d_B }{n_B^2} \cos(k_{Cz}d_C) \bigg)\bigg|_{k_\rho=0}.
	\end{split}
\end{equation}
\end{figure*}
Under the condition of the second type topological singularity, we can prove:
\begin{equation}
	(b-id)_s=(b-id)_p.
\end{equation}
Therefore, according to Eq. (\ref{Hps}), the transfer function can be obtained as:
\begin{equation}
	H\propto sin(2\varphi)k_\rho^4.
\end{equation}  
\nocite{*}

\bibliography{apssamp}

\providecommand{\noopsort}[1]{}\providecommand{\singleletter}[1]{#1}%
\begin{thebibliography}{41}%
\makeatletter
\providecommand \@ifxundefined [1]{%
 \@ifx{#1\undefined}
}%
\providecommand \@ifnum [1]{%
 \ifnum #1\expandafter \@firstoftwo
 \else \expandafter \@secondoftwo
 \fi
}%
\providecommand \@ifx [1]{%
 \ifx #1\expandafter \@firstoftwo
 \else \expandafter \@secondoftwo
 \fi
}%
\providecommand \natexlab [1]{#1}%
\providecommand \enquote  [1]{``#1''}%
\providecommand \bibnamefont  [1]{#1}%
\providecommand \bibfnamefont [1]{#1}%
\providecommand \citenamefont [1]{#1}%
\providecommand \href@noop [0]{\@secondoftwo}%
\providecommand \href [0]{\begingroup \@sanitize@url \@href}%
\providecommand \@href[1]{\@@startlink{#1}\@@href}%
\providecommand \@@href[1]{\endgroup#1\@@endlink}%
\providecommand \@sanitize@url [0]{\catcode `\\12\catcode `\$12\catcode
  `\&12\catcode `\#12\catcode `\^12\catcode `\_12\catcode `\%12\relax}%
\providecommand \@@startlink[1]{}%
\providecommand \@@endlink[0]{}%
\providecommand \url  [0]{\begingroup\@sanitize@url \@url }%
\providecommand \@url [1]{\endgroup\@href {#1}{\urlprefix }}%
\providecommand \urlprefix  [0]{URL }%
\providecommand \Eprint [0]{\href }%
\providecommand \doibase [0]{https://doi.org/}%
\providecommand \selectlanguage [0]{\@gobble}%
\providecommand \bibinfo  [0]{\@secondoftwo}%
\providecommand \bibfield  [0]{\@secondoftwo}%
\providecommand \translation [1]{[#1]}%
\providecommand \BibitemOpen [0]{}%
\providecommand \bibitemStop [0]{}%
\providecommand \bibitemNoStop [0]{.\EOS\space}%
\providecommand \EOS [0]{\spacefactor3000\relax}%
\providecommand \BibitemShut  [1]{\csname bibitem#1\endcsname}%
\let\auto@bib@innerbib\@empty
\bibitem [{\citenamefont {Li}\ \emph {et~al.}(2021)\citenamefont {Li},
  \citenamefont {Zhang}, \citenamefont {Li}, \citenamefont {Fang},\ and\
  \citenamefont {Dong}}]{li2021challenges}%
  \BibitemOpen
  \bibfield  {author} {\bibinfo {author} {\bibfnamefont {C.}~\bibnamefont
  {Li}}, \bibinfo {author} {\bibfnamefont {X.}~\bibnamefont {Zhang}}, \bibinfo
  {author} {\bibfnamefont {J.}~\bibnamefont {Li}}, \bibinfo {author}
  {\bibfnamefont {T.}~\bibnamefont {Fang}},\ and\ \bibinfo {author}
  {\bibfnamefont {X.}~\bibnamefont {Dong}},\ }\bibfield  {title} {\bibinfo
  {title} {The challenges of modern computing and new opportunities for
  optics},\ }\href@noop {} {\bibfield  {journal} {\bibinfo  {journal}
  {PhotoniX}\ }\textbf {\bibinfo {volume} {2}},\ \bibinfo {pages} {1} (\bibinfo
  {year} {2021})}\BibitemShut {NoStop}%
\bibitem [{\citenamefont {Wang}\ \emph {et~al.}(2022)\citenamefont {Wang},
  \citenamefont {Hu}, \citenamefont {Wang}, \citenamefont {Ding}, \citenamefont
  {Zhang}, \citenamefont {Li}, \citenamefont {Burokur}, \citenamefont {Wu},
  \citenamefont {Liu}, \citenamefont {Tan} \emph {et~al.}}]{wang2022single}%
  \BibitemOpen
  \bibfield  {author} {\bibinfo {author} {\bibfnamefont {Z.}~\bibnamefont
  {Wang}}, \bibinfo {author} {\bibfnamefont {G.}~\bibnamefont {Hu}}, \bibinfo
  {author} {\bibfnamefont {X.}~\bibnamefont {Wang}}, \bibinfo {author}
  {\bibfnamefont {X.}~\bibnamefont {Ding}}, \bibinfo {author} {\bibfnamefont
  {K.}~\bibnamefont {Zhang}}, \bibinfo {author} {\bibfnamefont
  {H.}~\bibnamefont {Li}}, \bibinfo {author} {\bibfnamefont {S.~N.}\
  \bibnamefont {Burokur}}, \bibinfo {author} {\bibfnamefont {Q.}~\bibnamefont
  {Wu}}, \bibinfo {author} {\bibfnamefont {J.}~\bibnamefont {Liu}}, \bibinfo
  {author} {\bibfnamefont {J.}~\bibnamefont {Tan}}, \emph {et~al.},\ }\bibfield
   {title} {\bibinfo {title} {Single-layer spatial analog meta-processor for
  imaging processing},\ }\href@noop {} {\bibfield  {journal} {\bibinfo
  {journal} {Nature communications}\ }\textbf {\bibinfo {volume} {13}},\
  \bibinfo {pages} {2188} (\bibinfo {year} {2022})}\BibitemShut {NoStop}%
\bibitem [{\citenamefont {Fu}\ \emph {et~al.}(2022)\citenamefont {Fu},
  \citenamefont {Zhao}, \citenamefont {Li}, \citenamefont {Liu}, \citenamefont
  {Tian},\ and\ \citenamefont {Huang}}]{fu2022ultracompact}%
  \BibitemOpen
  \bibfield  {author} {\bibinfo {author} {\bibfnamefont {W.}~\bibnamefont
  {Fu}}, \bibinfo {author} {\bibfnamefont {D.}~\bibnamefont {Zhao}}, \bibinfo
  {author} {\bibfnamefont {Z.}~\bibnamefont {Li}}, \bibinfo {author}
  {\bibfnamefont {S.}~\bibnamefont {Liu}}, \bibinfo {author} {\bibfnamefont
  {C.}~\bibnamefont {Tian}},\ and\ \bibinfo {author} {\bibfnamefont
  {K.}~\bibnamefont {Huang}},\ }\bibfield  {title} {\bibinfo {title}
  {Ultracompact meta-imagers for arbitrary all-optical convolution},\
  }\href@noop {} {\bibfield  {journal} {\bibinfo  {journal} {Light: Science \&
  Applications}\ }\textbf {\bibinfo {volume} {11}},\ \bibinfo {pages} {62}
  (\bibinfo {year} {2022})}\BibitemShut {NoStop}%
\bibitem [{\citenamefont {Wan}\ \emph {et~al.}(2021)\citenamefont {Wan},
  \citenamefont {Pan}, \citenamefont {Feng}, \citenamefont {Liu},\ and\
  \citenamefont {Potapov}}]{wan2021review}%
  \BibitemOpen
  \bibfield  {author} {\bibinfo {author} {\bibfnamefont {L.}~\bibnamefont
  {Wan}}, \bibinfo {author} {\bibfnamefont {D.}~\bibnamefont {Pan}}, \bibinfo
  {author} {\bibfnamefont {T.}~\bibnamefont {Feng}}, \bibinfo {author}
  {\bibfnamefont {W.}~\bibnamefont {Liu}},\ and\ \bibinfo {author}
  {\bibfnamefont {A.~A.}\ \bibnamefont {Potapov}},\ }\bibfield  {title}
  {\bibinfo {title} {A review of dielectric optical metasurfaces for spatial
  differentiation and edge detection},\ }\href@noop {} {\bibfield  {journal}
  {\bibinfo  {journal} {Frontiers of Optoelectronics}\ }\textbf {\bibinfo
  {volume} {14}},\ \bibinfo {pages} {187} (\bibinfo {year} {2021})}\BibitemShut
  {NoStop}%
\bibitem [{\citenamefont {Tang}\ \emph {et~al.}(2000)\citenamefont {Tang},
  \citenamefont {Wu}, \citenamefont {Ma}, \citenamefont {Gallagher},
  \citenamefont {Perera},\ and\ \citenamefont {Zhuang}}]{tang2000mri}%
  \BibitemOpen
  \bibfield  {author} {\bibinfo {author} {\bibfnamefont {H.}~\bibnamefont
  {Tang}}, \bibinfo {author} {\bibfnamefont {E.}~\bibnamefont {Wu}}, \bibinfo
  {author} {\bibfnamefont {Q.}~\bibnamefont {Ma}}, \bibinfo {author}
  {\bibfnamefont {D.}~\bibnamefont {Gallagher}}, \bibinfo {author}
  {\bibfnamefont {G.}~\bibnamefont {Perera}},\ and\ \bibinfo {author}
  {\bibfnamefont {T.}~\bibnamefont {Zhuang}},\ }\bibfield  {title} {\bibinfo
  {title} {Mri brain image segmentation by multi-resolution edge detection and
  region selection},\ }\href@noop {} {\bibfield  {journal} {\bibinfo  {journal}
  {Computerized Medical Imaging and Graphics}\ }\textbf {\bibinfo {volume}
  {24}},\ \bibinfo {pages} {349} (\bibinfo {year} {2000})}\BibitemShut
  {NoStop}%
\bibitem [{\citenamefont {Rajab}\ \emph {et~al.}(2004)\citenamefont {Rajab},
  \citenamefont {Woolfson},\ and\ \citenamefont
  {Morgan}}]{rajab2004application}%
  \BibitemOpen
  \bibfield  {author} {\bibinfo {author} {\bibfnamefont {M.}~\bibnamefont
  {Rajab}}, \bibinfo {author} {\bibfnamefont {M.}~\bibnamefont {Woolfson}},\
  and\ \bibinfo {author} {\bibfnamefont {S.}~\bibnamefont {Morgan}},\
  }\bibfield  {title} {\bibinfo {title} {Application of region-based
  segmentation and neural network edge detection to skin lesions},\ }\href@noop
  {} {\bibfield  {journal} {\bibinfo  {journal} {Computerized Medical Imaging
  and Graphics}\ }\textbf {\bibinfo {volume} {28}},\ \bibinfo {pages} {61}
  (\bibinfo {year} {2004})}\BibitemShut {NoStop}%
\bibitem [{\citenamefont {Zhu}\ \emph {et~al.}(2019)\citenamefont {Zhu},
  \citenamefont {Lou}, \citenamefont {Zhou}, \citenamefont {Zhang},
  \citenamefont {Huang}, \citenamefont {Li}, \citenamefont {Luo}, \citenamefont
  {Wen}, \citenamefont {Zhu}, \citenamefont {Gong} \emph
  {et~al.}}]{zhu2019generalized}%
  \BibitemOpen
  \bibfield  {author} {\bibinfo {author} {\bibfnamefont {T.}~\bibnamefont
  {Zhu}}, \bibinfo {author} {\bibfnamefont {Y.}~\bibnamefont {Lou}}, \bibinfo
  {author} {\bibfnamefont {Y.}~\bibnamefont {Zhou}}, \bibinfo {author}
  {\bibfnamefont {J.}~\bibnamefont {Zhang}}, \bibinfo {author} {\bibfnamefont
  {J.}~\bibnamefont {Huang}}, \bibinfo {author} {\bibfnamefont
  {Y.}~\bibnamefont {Li}}, \bibinfo {author} {\bibfnamefont {H.}~\bibnamefont
  {Luo}}, \bibinfo {author} {\bibfnamefont {S.}~\bibnamefont {Wen}}, \bibinfo
  {author} {\bibfnamefont {S.}~\bibnamefont {Zhu}}, \bibinfo {author}
  {\bibfnamefont {Q.}~\bibnamefont {Gong}}, \emph {et~al.},\ }\bibfield
  {title} {\bibinfo {title} {Generalized spatial differentiation from the spin
  hall effect of light and its application in image processing of edge
  detection},\ }\href@noop {} {\bibfield  {journal} {\bibinfo  {journal}
  {Physical Review Applied}\ }\textbf {\bibinfo {volume} {11}},\ \bibinfo
  {pages} {034043} (\bibinfo {year} {2019})}\BibitemShut {NoStop}%
\bibitem [{\citenamefont {Xu}\ \emph {et~al.}(2020{\natexlab{a}})\citenamefont
  {Xu}, \citenamefont {He}, \citenamefont {Zhou}, \citenamefont {Chen},
  \citenamefont {Wen},\ and\ \citenamefont {Luo}}]{xu2020optical}%
  \BibitemOpen
  \bibfield  {author} {\bibinfo {author} {\bibfnamefont {D.}~\bibnamefont
  {Xu}}, \bibinfo {author} {\bibfnamefont {S.}~\bibnamefont {He}}, \bibinfo
  {author} {\bibfnamefont {J.}~\bibnamefont {Zhou}}, \bibinfo {author}
  {\bibfnamefont {S.}~\bibnamefont {Chen}}, \bibinfo {author} {\bibfnamefont
  {S.}~\bibnamefont {Wen}},\ and\ \bibinfo {author} {\bibfnamefont
  {H.}~\bibnamefont {Luo}},\ }\bibfield  {title} {\bibinfo {title} {Optical
  analog computing of two-dimensional spatial differentiation based on the
  brewster effect},\ }\href@noop {} {\bibfield  {journal} {\bibinfo  {journal}
  {Optics Letters}\ }\textbf {\bibinfo {volume} {45}},\ \bibinfo {pages} {6867}
  (\bibinfo {year} {2020}{\natexlab{a}})}\BibitemShut {NoStop}%
\bibitem [{\citenamefont {Xu}\ \emph {et~al.}(2021)\citenamefont {Xu},
  \citenamefont {Ling}, \citenamefont {Xu}, \citenamefont {Chen}, \citenamefont
  {Wen},\ and\ \citenamefont {Luo}}]{xu2021enhanced}%
  \BibitemOpen
  \bibfield  {author} {\bibinfo {author} {\bibfnamefont {W.}~\bibnamefont
  {Xu}}, \bibinfo {author} {\bibfnamefont {X.}~\bibnamefont {Ling}}, \bibinfo
  {author} {\bibfnamefont {D.}~\bibnamefont {Xu}}, \bibinfo {author}
  {\bibfnamefont {S.}~\bibnamefont {Chen}}, \bibinfo {author} {\bibfnamefont
  {S.}~\bibnamefont {Wen}},\ and\ \bibinfo {author} {\bibfnamefont
  {H.}~\bibnamefont {Luo}},\ }\bibfield  {title} {\bibinfo {title} {Enhanced
  optical spatial differential operations via strong spin-orbit interactions in
  an anisotropic epsilon-near-zero slab},\ }\href@noop {} {\bibfield  {journal}
  {\bibinfo  {journal} {Physical Review A}\ }\textbf {\bibinfo {volume}
  {104}},\ \bibinfo {pages} {053513} (\bibinfo {year} {2021})}\BibitemShut
  {NoStop}%
\bibitem [{\citenamefont {Xia}\ \emph {et~al.}(2021)\citenamefont {Xia},
  \citenamefont {Wang},\ and\ \citenamefont {Zhi}}]{xia2021tunable}%
  \BibitemOpen
  \bibfield  {author} {\bibinfo {author} {\bibfnamefont {D.}~\bibnamefont
  {Xia}}, \bibinfo {author} {\bibfnamefont {Y.}~\bibnamefont {Wang}},\ and\
  \bibinfo {author} {\bibfnamefont {Q.}~\bibnamefont {Zhi}},\ }\bibfield
  {title} {\bibinfo {title} {Tunable optical differential operation based on
  the cross-polarization effect at the optical interface},\ }\href@noop {}
  {\bibfield  {journal} {\bibinfo  {journal} {Optics Express}\ }\textbf
  {\bibinfo {volume} {29}},\ \bibinfo {pages} {31891} (\bibinfo {year}
  {2021})}\BibitemShut {NoStop}%
\bibitem [{\citenamefont {Mi}\ \emph {et~al.}(2020)\citenamefont {Mi},
  \citenamefont {Song}, \citenamefont {Cai}, \citenamefont {Yang},
  \citenamefont {Song},\ and\ \citenamefont {Mi}}]{mi2020tunable}%
  \BibitemOpen
  \bibfield  {author} {\bibinfo {author} {\bibfnamefont {C.}~\bibnamefont
  {Mi}}, \bibinfo {author} {\bibfnamefont {W.}~\bibnamefont {Song}}, \bibinfo
  {author} {\bibfnamefont {X.}~\bibnamefont {Cai}}, \bibinfo {author}
  {\bibfnamefont {C.}~\bibnamefont {Yang}}, \bibinfo {author} {\bibfnamefont
  {Y.}~\bibnamefont {Song}},\ and\ \bibinfo {author} {\bibfnamefont
  {X.}~\bibnamefont {Mi}},\ }\bibfield  {title} {\bibinfo {title} {Tunable
  optical spatial differentiation in the photonic spin hall effect},\
  }\href@noop {} {\bibfield  {journal} {\bibinfo  {journal} {Optics Express}\
  }\textbf {\bibinfo {volume} {28}},\ \bibinfo {pages} {30222} (\bibinfo {year}
  {2020})}\BibitemShut {NoStop}%
\bibitem [{\citenamefont {Xu}\ \emph {et~al.}(2020{\natexlab{b}})\citenamefont
  {Xu}, \citenamefont {He}, \citenamefont {Zhou}, \citenamefont {Chen},
  \citenamefont {Wen},\ and\ \citenamefont {Luo}}]{xu2020goos}%
  \BibitemOpen
  \bibfield  {author} {\bibinfo {author} {\bibfnamefont {D.}~\bibnamefont
  {Xu}}, \bibinfo {author} {\bibfnamefont {S.}~\bibnamefont {He}}, \bibinfo
  {author} {\bibfnamefont {J.}~\bibnamefont {Zhou}}, \bibinfo {author}
  {\bibfnamefont {S.}~\bibnamefont {Chen}}, \bibinfo {author} {\bibfnamefont
  {S.}~\bibnamefont {Wen}},\ and\ \bibinfo {author} {\bibfnamefont
  {H.}~\bibnamefont {Luo}},\ }\bibfield  {title} {\bibinfo {title}
  {Goos--h{\"a}nchen effect enabled optical differential operation and image
  edge detection},\ }\href@noop {} {\bibfield  {journal} {\bibinfo  {journal}
  {Applied Physics Letters}\ }\textbf {\bibinfo {volume} {116}} (\bibinfo
  {year} {2020}{\natexlab{b}})}\BibitemShut {NoStop}%
\bibitem [{\citenamefont {Deng}\ \emph
  {et~al.}(2024{\natexlab{a}})\citenamefont {Deng}, \citenamefont {Xu},
  \citenamefont {Zhang}, \citenamefont {Yang}, \citenamefont {Xu},\ and\
  \citenamefont {Luo}}]{deng2024r}%
  \BibitemOpen
  \bibfield  {author} {\bibinfo {author} {\bibfnamefont {Y.}~\bibnamefont
  {Deng}}, \bibinfo {author} {\bibfnamefont {W.}~\bibnamefont {Xu}}, \bibinfo
  {author} {\bibfnamefont {W.}~\bibnamefont {Zhang}}, \bibinfo {author}
  {\bibfnamefont {Q.}~\bibnamefont {Yang}}, \bibinfo {author} {\bibfnamefont
  {D.}~\bibnamefont {Xu}},\ and\ \bibinfo {author} {\bibfnamefont
  {H.}~\bibnamefont {Luo}},\ }\bibfield  {title} {\bibinfo {title} {Rotational
  photonic spin hall effect on twisted bilayer metasurfaces},\ }\href@noop {}
  {\bibfield  {journal} {\bibinfo  {journal} {Optics Communications}\ }\textbf
  {\bibinfo {volume} {560}},\ \bibinfo {pages} {130480} (\bibinfo {year}
  {2024}{\natexlab{a}})}\BibitemShut {NoStop}%
\bibitem [{\citenamefont {Wen}\ \emph {et~al.}(2024)\citenamefont {Wen},
  \citenamefont {Xu}, \citenamefont {Zhang}, \citenamefont {Jiang},\ and\
  \citenamefont {Luo}}]{wen2024tunable}%
  \BibitemOpen
  \bibfield  {author} {\bibinfo {author} {\bibfnamefont {Z.}~\bibnamefont
  {Wen}}, \bibinfo {author} {\bibfnamefont {W.}~\bibnamefont {Xu}}, \bibinfo
  {author} {\bibfnamefont {Y.}~\bibnamefont {Zhang}}, \bibinfo {author}
  {\bibfnamefont {T.}~\bibnamefont {Jiang}},\ and\ \bibinfo {author}
  {\bibfnamefont {Z.}~\bibnamefont {Luo}},\ }\bibfield  {title} {\bibinfo
  {title} {Tunable optical spatial differential operation via photonic spin
  hall effect in a weyl semimetal},\ }\href@noop {} {\bibfield  {journal}
  {\bibinfo  {journal} {Optics Express}\ }\textbf {\bibinfo {volume} {32}},\
  \bibinfo {pages} {10022} (\bibinfo {year} {2024})}\BibitemShut {NoStop}%
\bibitem [{\citenamefont {Guo}\ \emph {et~al.}(2018)\citenamefont {Guo},
  \citenamefont {Xiao}, \citenamefont {Minkov}, \citenamefont {Shi},\ and\
  \citenamefont {Fan}}]{guo2018photonic}%
  \BibitemOpen
  \bibfield  {author} {\bibinfo {author} {\bibfnamefont {C.}~\bibnamefont
  {Guo}}, \bibinfo {author} {\bibfnamefont {M.}~\bibnamefont {Xiao}}, \bibinfo
  {author} {\bibfnamefont {M.}~\bibnamefont {Minkov}}, \bibinfo {author}
  {\bibfnamefont {Y.}~\bibnamefont {Shi}},\ and\ \bibinfo {author}
  {\bibfnamefont {S.}~\bibnamefont {Fan}},\ }\bibfield  {title} {\bibinfo
  {title} {Photonic crystal slab laplace operator for image differentiation},\
  }\href@noop {} {\bibfield  {journal} {\bibinfo  {journal} {Optica}\ }\textbf
  {\bibinfo {volume} {5}},\ \bibinfo {pages} {251} (\bibinfo {year}
  {2018})}\BibitemShut {NoStop}%
\bibitem [{\citenamefont {Zhu}\ \emph {et~al.}(2021)\citenamefont {Zhu},
  \citenamefont {Guo}, \citenamefont {Huang}, \citenamefont {Wang},
  \citenamefont {Orenstein}, \citenamefont {Ruan},\ and\ \citenamefont
  {Fan}}]{zhu2021topological}%
  \BibitemOpen
  \bibfield  {author} {\bibinfo {author} {\bibfnamefont {T.}~\bibnamefont
  {Zhu}}, \bibinfo {author} {\bibfnamefont {C.}~\bibnamefont {Guo}}, \bibinfo
  {author} {\bibfnamefont {J.}~\bibnamefont {Huang}}, \bibinfo {author}
  {\bibfnamefont {H.}~\bibnamefont {Wang}}, \bibinfo {author} {\bibfnamefont
  {M.}~\bibnamefont {Orenstein}}, \bibinfo {author} {\bibfnamefont
  {Z.}~\bibnamefont {Ruan}},\ and\ \bibinfo {author} {\bibfnamefont
  {S.}~\bibnamefont {Fan}},\ }\bibfield  {title} {\bibinfo {title} {Topological
  optical differentiator},\ }\href@noop {} {\bibfield  {journal} {\bibinfo
  {journal} {Nature communications}\ }\textbf {\bibinfo {volume} {12}},\
  \bibinfo {pages} {680} (\bibinfo {year} {2021})}\BibitemShut {NoStop}%
\bibitem [{\citenamefont {Zhou}\ \emph {et~al.}(2020)\citenamefont {Zhou},
  \citenamefont {Zheng}, \citenamefont {Kravchenko},\ and\ \citenamefont
  {Valentine}}]{zhou2020flat}%
  \BibitemOpen
  \bibfield  {author} {\bibinfo {author} {\bibfnamefont {Y.}~\bibnamefont
  {Zhou}}, \bibinfo {author} {\bibfnamefont {H.}~\bibnamefont {Zheng}},
  \bibinfo {author} {\bibfnamefont {I.~I.}\ \bibnamefont {Kravchenko}},\ and\
  \bibinfo {author} {\bibfnamefont {J.}~\bibnamefont {Valentine}},\ }\bibfield
  {title} {\bibinfo {title} {Flat optics for image differentiation},\
  }\href@noop {} {\bibfield  {journal} {\bibinfo  {journal} {Nature Photonics}\
  }\textbf {\bibinfo {volume} {14}},\ \bibinfo {pages} {316} (\bibinfo {year}
  {2020})}\BibitemShut {NoStop}%
\bibitem [{\citenamefont {Zhang}\ \emph {et~al.}(2017)\citenamefont {Zhang},
  \citenamefont {Pu}, \citenamefont {Li}, \citenamefont {Gao}, \citenamefont
  {Ma}, \citenamefont {Luo}, \citenamefont {Yu},\ and\ \citenamefont
  {Luo}}]{zhang2017all}%
  \BibitemOpen
  \bibfield  {author} {\bibinfo {author} {\bibfnamefont {F.}~\bibnamefont
  {Zhang}}, \bibinfo {author} {\bibfnamefont {M.}~\bibnamefont {Pu}}, \bibinfo
  {author} {\bibfnamefont {X.}~\bibnamefont {Li}}, \bibinfo {author}
  {\bibfnamefont {P.}~\bibnamefont {Gao}}, \bibinfo {author} {\bibfnamefont
  {X.}~\bibnamefont {Ma}}, \bibinfo {author} {\bibfnamefont {J.}~\bibnamefont
  {Luo}}, \bibinfo {author} {\bibfnamefont {H.}~\bibnamefont {Yu}},\ and\
  \bibinfo {author} {\bibfnamefont {X.}~\bibnamefont {Luo}},\ }\bibfield
  {title} {\bibinfo {title} {All-dielectric metasurfaces for simultaneous giant
  circular asymmetric transmission and wavefront shaping based on asymmetric
  photonic spin--orbit interactions},\ }\href@noop {} {\bibfield  {journal}
  {\bibinfo  {journal} {Advanced Functional Materials}\ }\textbf {\bibinfo
  {volume} {27}},\ \bibinfo {pages} {1704295} (\bibinfo {year}
  {2017})}\BibitemShut {NoStop}%
\bibitem [{\citenamefont {Pan}\ \emph {et~al.}(2021)\citenamefont {Pan},
  \citenamefont {Wan}, \citenamefont {Ouyang}, \citenamefont {Zhang},
  \citenamefont {Potapov}, \citenamefont {Liu}, \citenamefont {Liang},
  \citenamefont {Feng},\ and\ \citenamefont {Li}}]{pan2021laplace}%
  \BibitemOpen
  \bibfield  {author} {\bibinfo {author} {\bibfnamefont {D.}~\bibnamefont
  {Pan}}, \bibinfo {author} {\bibfnamefont {L.}~\bibnamefont {Wan}}, \bibinfo
  {author} {\bibfnamefont {M.}~\bibnamefont {Ouyang}}, \bibinfo {author}
  {\bibfnamefont {W.}~\bibnamefont {Zhang}}, \bibinfo {author} {\bibfnamefont
  {A.~A.}\ \bibnamefont {Potapov}}, \bibinfo {author} {\bibfnamefont
  {W.}~\bibnamefont {Liu}}, \bibinfo {author} {\bibfnamefont {Z.}~\bibnamefont
  {Liang}}, \bibinfo {author} {\bibfnamefont {T.}~\bibnamefont {Feng}},\ and\
  \bibinfo {author} {\bibfnamefont {Z.}~\bibnamefont {Li}},\ }\bibfield
  {title} {\bibinfo {title} {Laplace metasurfaces for optical analog computing
  based on quasi-bound states in the continuum},\ }\href@noop {} {\bibfield
  {journal} {\bibinfo  {journal} {Photonics Research}\ }\textbf {\bibinfo
  {volume} {9}},\ \bibinfo {pages} {1758} (\bibinfo {year} {2021})}\BibitemShut
  {NoStop}%
\bibitem [{\citenamefont {Deng}\ \emph
  {et~al.}(2024{\natexlab{b}})\citenamefont {Deng}, \citenamefont {Cotrufo},
  \citenamefont {Wang}, \citenamefont {Dong}, \citenamefont {Ruan},
  \citenamefont {Al{\`u}},\ and\ \citenamefont {Chen}}]{deng2024broadband}%
  \BibitemOpen
  \bibfield  {author} {\bibinfo {author} {\bibfnamefont {M.}~\bibnamefont
  {Deng}}, \bibinfo {author} {\bibfnamefont {M.}~\bibnamefont {Cotrufo}},
  \bibinfo {author} {\bibfnamefont {J.}~\bibnamefont {Wang}}, \bibinfo {author}
  {\bibfnamefont {J.}~\bibnamefont {Dong}}, \bibinfo {author} {\bibfnamefont
  {Z.}~\bibnamefont {Ruan}}, \bibinfo {author} {\bibfnamefont {A.}~\bibnamefont
  {Al{\`u}}},\ and\ \bibinfo {author} {\bibfnamefont {L.}~\bibnamefont
  {Chen}},\ }\bibfield  {title} {\bibinfo {title} {Broadband angular spectrum
  differentiation using dielectric metasurfaces},\ }\href@noop {} {\bibfield
  {journal} {\bibinfo  {journal} {Nature Communications}\ }\textbf {\bibinfo
  {volume} {15}},\ \bibinfo {pages} {2237} (\bibinfo {year}
  {2024}{\natexlab{b}})}\BibitemShut {NoStop}%
\bibitem [{\citenamefont {Liu}\ \emph {et~al.}(2022)\citenamefont {Liu},
  \citenamefont {Huang}, \citenamefont {Chen},\ and\ \citenamefont
  {Zhang}}]{liu2022single}%
  \BibitemOpen
  \bibfield  {author} {\bibinfo {author} {\bibfnamefont {Y.}~\bibnamefont
  {Liu}}, \bibinfo {author} {\bibfnamefont {M.}~\bibnamefont {Huang}}, \bibinfo
  {author} {\bibfnamefont {Q.}~\bibnamefont {Chen}},\ and\ \bibinfo {author}
  {\bibfnamefont {D.}~\bibnamefont {Zhang}},\ }\bibfield  {title} {\bibinfo
  {title} {Single planar photonic chip with tailored angular transmission for
  multiple-order analog spatial differentiator},\ }\href@noop {} {\bibfield
  {journal} {\bibinfo  {journal} {Nature Communications}\ }\textbf {\bibinfo
  {volume} {13}},\ \bibinfo {pages} {7944} (\bibinfo {year}
  {2022})}\BibitemShut {NoStop}%
\bibitem [{\citenamefont {Hasan}\ and\ \citenamefont
  {Kane}(2010)}]{hasan2010colloquium}%
  \BibitemOpen
  \bibfield  {author} {\bibinfo {author} {\bibfnamefont {M.~Z.}\ \bibnamefont
  {Hasan}}\ and\ \bibinfo {author} {\bibfnamefont {C.~L.}\ \bibnamefont
  {Kane}},\ }\bibfield  {title} {\bibinfo {title} {Colloquium: topological
  insulators},\ }\href@noop {} {\bibfield  {journal} {\bibinfo  {journal}
  {Reviews of modern physics}\ }\textbf {\bibinfo {volume} {82}},\ \bibinfo
  {pages} {3045} (\bibinfo {year} {2010})}\BibitemShut {NoStop}%
\bibitem [{\citenamefont {Shen}(2012)}]{shen2012topological}%
  \BibitemOpen
  \bibfield  {author} {\bibinfo {author} {\bibfnamefont {S.-Q.}\ \bibnamefont
  {Shen}},\ }\href@noop {} {\emph {\bibinfo {title} {Topological
  insulators}}},\ Vol.\ \bibinfo {volume} {174}\ (\bibinfo  {publisher}
  {Springer},\ \bibinfo {year} {2012})\BibitemShut {NoStop}%
\bibitem [{\citenamefont {Lu}\ \emph {et~al.}(2014)\citenamefont {Lu},
  \citenamefont {Joannopoulos},\ and\ \citenamefont
  {Solja{\v{c}}i{\'c}}}]{lu2014topological}%
  \BibitemOpen
  \bibfield  {author} {\bibinfo {author} {\bibfnamefont {L.}~\bibnamefont
  {Lu}}, \bibinfo {author} {\bibfnamefont {J.~D.}\ \bibnamefont
  {Joannopoulos}},\ and\ \bibinfo {author} {\bibfnamefont {M.}~\bibnamefont
  {Solja{\v{c}}i{\'c}}},\ }\bibfield  {title} {\bibinfo {title} {Topological
  photonics},\ }\href@noop {} {\bibfield  {journal} {\bibinfo  {journal}
  {Nature photonics}\ }\textbf {\bibinfo {volume} {8}},\ \bibinfo {pages} {821}
  (\bibinfo {year} {2014})}\BibitemShut {NoStop}%
\bibitem [{\citenamefont {Xiao}\ \emph {et~al.}(2014)\citenamefont {Xiao},
  \citenamefont {Zhang},\ and\ \citenamefont {Chan}}]{xiao2014surface}%
  \BibitemOpen
  \bibfield  {author} {\bibinfo {author} {\bibfnamefont {M.}~\bibnamefont
  {Xiao}}, \bibinfo {author} {\bibfnamefont {Z.}~\bibnamefont {Zhang}},\ and\
  \bibinfo {author} {\bibfnamefont {C.~T.}\ \bibnamefont {Chan}},\ }\bibfield
  {title} {\bibinfo {title} {Surface impedance and bulk band geometric phases
  in one-dimensional systems},\ }\href@noop {} {\bibfield  {journal} {\bibinfo
  {journal} {Physical Review X}\ }\textbf {\bibinfo {volume} {4}},\ \bibinfo
  {pages} {021017} (\bibinfo {year} {2014})}\BibitemShut {NoStop}%
\bibitem [{\citenamefont {Bansil}\ \emph {et~al.}(2016)\citenamefont {Bansil},
  \citenamefont {Lin},\ and\ \citenamefont {Das}}]{bansil2016colloquium}%
  \BibitemOpen
  \bibfield  {author} {\bibinfo {author} {\bibfnamefont {A.}~\bibnamefont
  {Bansil}}, \bibinfo {author} {\bibfnamefont {H.}~\bibnamefont {Lin}},\ and\
  \bibinfo {author} {\bibfnamefont {T.}~\bibnamefont {Das}},\ }\bibfield
  {title} {\bibinfo {title} {Colloquium: Topological band theory},\ }\href@noop
  {} {\bibfield  {journal} {\bibinfo  {journal} {Reviews of Modern Physics}\
  }\textbf {\bibinfo {volume} {88}},\ \bibinfo {pages} {021004} (\bibinfo
  {year} {2016})}\BibitemShut {NoStop}%
\bibitem [{\citenamefont {Chiu}\ \emph {et~al.}(2016)\citenamefont {Chiu},
  \citenamefont {Teo}, \citenamefont {Schnyder},\ and\ \citenamefont
  {Ryu}}]{chiu2016classification}%
  \BibitemOpen
  \bibfield  {author} {\bibinfo {author} {\bibfnamefont {C.-K.}\ \bibnamefont
  {Chiu}}, \bibinfo {author} {\bibfnamefont {J.~C.}\ \bibnamefont {Teo}},
  \bibinfo {author} {\bibfnamefont {A.~P.}\ \bibnamefont {Schnyder}},\ and\
  \bibinfo {author} {\bibfnamefont {S.}~\bibnamefont {Ryu}},\ }\bibfield
  {title} {\bibinfo {title} {Classification of topological quantum matter with
  symmetries},\ }\href@noop {} {\bibfield  {journal} {\bibinfo  {journal}
  {Reviews of Modern Physics}\ }\textbf {\bibinfo {volume} {88}},\ \bibinfo
  {pages} {035005} (\bibinfo {year} {2016})}\BibitemShut {NoStop}%
\bibitem [{\citenamefont {Khanikaev}\ and\ \citenamefont
  {Shvets}(2017)}]{khanikaev2017two}%
  \BibitemOpen
  \bibfield  {author} {\bibinfo {author} {\bibfnamefont {A.~B.}\ \bibnamefont
  {Khanikaev}}\ and\ \bibinfo {author} {\bibfnamefont {G.}~\bibnamefont
  {Shvets}},\ }\bibfield  {title} {\bibinfo {title} {Two-dimensional
  topological photonics},\ }\href@noop {} {\bibfield  {journal} {\bibinfo
  {journal} {Nature photonics}\ }\textbf {\bibinfo {volume} {11}},\ \bibinfo
  {pages} {763} (\bibinfo {year} {2017})}\BibitemShut {NoStop}%
\bibitem [{\citenamefont {Wang}\ \emph {et~al.}(2020)\citenamefont {Wang},
  \citenamefont {Gupta}, \citenamefont {Xie},\ and\ \citenamefont
  {Lu}}]{wang2020topological}%
  \BibitemOpen
  \bibfield  {author} {\bibinfo {author} {\bibfnamefont {H.}~\bibnamefont
  {Wang}}, \bibinfo {author} {\bibfnamefont {S.~K.}\ \bibnamefont {Gupta}},
  \bibinfo {author} {\bibfnamefont {B.}~\bibnamefont {Xie}},\ and\ \bibinfo
  {author} {\bibfnamefont {M.}~\bibnamefont {Lu}},\ }\bibfield  {title}
  {\bibinfo {title} {Topological photonic crystals: a review},\ }\href@noop {}
  {\bibfield  {journal} {\bibinfo  {journal} {Frontiers of Optoelectronics}\
  }\textbf {\bibinfo {volume} {13}},\ \bibinfo {pages} {50} (\bibinfo {year}
  {2020})}\BibitemShut {NoStop}%
\bibitem [{\citenamefont {Xiong}\ \emph {et~al.}(2021)\citenamefont {Xiong},
  \citenamefont {Zhang},\ and\ \citenamefont {Jiang}}]{xiong2021resonance}%
  \BibitemOpen
  \bibfield  {author} {\bibinfo {author} {\bibfnamefont {L.}~\bibnamefont
  {Xiong}}, \bibinfo {author} {\bibfnamefont {Y.}~\bibnamefont {Zhang}},\ and\
  \bibinfo {author} {\bibfnamefont {X.}~\bibnamefont {Jiang}},\ }\bibfield
  {title} {\bibinfo {title} {Resonance and topological singularity near and
  beyond zero frequency for waves: model, theory, and effects},\ }\href@noop {}
  {\bibfield  {journal} {\bibinfo  {journal} {Photonics Research}\ }\textbf
  {\bibinfo {volume} {9}},\ \bibinfo {pages} {2024} (\bibinfo {year}
  {2021})}\BibitemShut {NoStop}%
\bibitem [{\citenamefont {Asb{\'o}th}\ and\ \citenamefont
  {Obuse}(2013)}]{asboth2013bulk}%
  \BibitemOpen
  \bibfield  {author} {\bibinfo {author} {\bibfnamefont {J.~K.}\ \bibnamefont
  {Asb{\'o}th}}\ and\ \bibinfo {author} {\bibfnamefont {H.}~\bibnamefont
  {Obuse}},\ }\bibfield  {title} {\bibinfo {title} {Bulk-boundary
  correspondence for chiral symmetric quantum walks},\ }\href@noop {}
  {\bibfield  {journal} {\bibinfo  {journal} {Physical Review B—Condensed
  Matter and Materials Physics}\ }\textbf {\bibinfo {volume} {88}},\ \bibinfo
  {pages} {121406} (\bibinfo {year} {2013})}\BibitemShut {NoStop}%
\bibitem [{\citenamefont {Batra}\ and\ \citenamefont
  {Sheet}(2019)}]{batra2019understanding}%
  \BibitemOpen
  \bibfield  {author} {\bibinfo {author} {\bibfnamefont {N.}~\bibnamefont
  {Batra}}\ and\ \bibinfo {author} {\bibfnamefont {G.}~\bibnamefont {Sheet}},\
  }\bibfield  {title} {\bibinfo {title} {Understanding basic concepts of
  topological insulators through su-schrieffer-heeger (ssh) model},\
  }\href@noop {} {\bibfield  {journal} {\bibinfo  {journal} {arXiv preprint
  arXiv:1906.08435}\ } (\bibinfo {year} {2019})}\BibitemShut {NoStop}%
\bibitem [{\citenamefont {Li}\ \emph {et~al.}(2019)\citenamefont {Li},
  \citenamefont {Zhang},\ and\ \citenamefont {Jiang}}]{li2019two}%
  \BibitemOpen
  \bibfield  {author} {\bibinfo {author} {\bibfnamefont {Q.}~\bibnamefont
  {Li}}, \bibinfo {author} {\bibfnamefont {Y.}~\bibnamefont {Zhang}},\ and\
  \bibinfo {author} {\bibfnamefont {X.}~\bibnamefont {Jiang}},\ }\bibfield
  {title} {\bibinfo {title} {Two classes of singularities and novel topology in
  a specially designed synthetic photonic crystals},\ }\href@noop {} {\bibfield
   {journal} {\bibinfo  {journal} {Optics express}\ }\textbf {\bibinfo {volume}
  {27}},\ \bibinfo {pages} {4956} (\bibinfo {year} {2019})}\BibitemShut
  {NoStop}%
\bibitem [{\citenamefont {Li}\ and\ \citenamefont
  {Jiang}(2019)}]{li2019singularity}%
  \BibitemOpen
  \bibfield  {author} {\bibinfo {author} {\bibfnamefont {Q.}~\bibnamefont
  {Li}}\ and\ \bibinfo {author} {\bibfnamefont {X.}~\bibnamefont {Jiang}},\
  }\bibfield  {title} {\bibinfo {title} {Singularity induced topological
  transition of different dimensions in one synthetic photonic system},\
  }\href@noop {} {\bibfield  {journal} {\bibinfo  {journal} {Optics
  Communications}\ }\textbf {\bibinfo {volume} {440}},\ \bibinfo {pages} {32}
  (\bibinfo {year} {2019})}\BibitemShut {NoStop}%
\bibitem [{\citenamefont {Liu}\ \emph {et~al.}(2023)\citenamefont {Liu},
  \citenamefont {Xiong},\ and\ \citenamefont {Jiang}}]{liu2023evolution}%
  \BibitemOpen
  \bibfield  {author} {\bibinfo {author} {\bibfnamefont {Y.}~\bibnamefont
  {Liu}}, \bibinfo {author} {\bibfnamefont {L.}~\bibnamefont {Xiong}},\ and\
  \bibinfo {author} {\bibfnamefont {X.}~\bibnamefont {Jiang}},\ }\bibfield
  {title} {\bibinfo {title} {The evolution of topological singularities between
  real-and complex-frequency domains and the engineering of photonic bands for
  hermitian and non-hermitian photonic crystals},\ }\href@noop {} {\bibfield
  {journal} {\bibinfo  {journal} {New Journal of Physics}\ }\textbf {\bibinfo
  {volume} {24}},\ \bibinfo {pages} {123042} (\bibinfo {year}
  {2023})}\BibitemShut {NoStop}%
\bibitem [{\citenamefont {Liu}\ \emph {et~al.}()\citenamefont {Liu},
  \citenamefont {Wang}, \citenamefont {Li}, \citenamefont {Zhang},
  \citenamefont {Wang}, \citenamefont {Lai},\ and\ \citenamefont
  {Jiang}}]{liudual}%
  \BibitemOpen
  \bibfield  {author} {\bibinfo {author} {\bibfnamefont {Y.}~\bibnamefont
  {Liu}}, \bibinfo {author} {\bibfnamefont {X.}~\bibnamefont {Wang}}, \bibinfo
  {author} {\bibfnamefont {Y.}~\bibnamefont {Li}}, \bibinfo {author}
  {\bibfnamefont {H.}~\bibnamefont {Zhang}}, \bibinfo {author} {\bibfnamefont
  {X.}~\bibnamefont {Wang}}, \bibinfo {author} {\bibfnamefont {Z.}~\bibnamefont
  {Lai}},\ and\ \bibinfo {author} {\bibfnamefont {X.}~\bibnamefont {Jiang}},\
  }\bibfield  {title} {\bibinfo {title} {Dual-polarization huge photonic spin
  hall shift and super-subwavelength detecting based on topological
  singularities in 1d photonic crystals},\ }\href@noop {} {\bibinfo  {journal}
  {Laser \& Photonics Reviews}\ ,\ \bibinfo {pages} {2400973}}\BibitemShut
  {NoStop}%
\bibitem [{\citenamefont {Shou}\ \emph {et~al.}(2022)\citenamefont {Shou},
  \citenamefont {Wang}, \citenamefont {Miao}, \citenamefont {Chen},\ and\
  \citenamefont {Luo}}]{shou2022realization}%
  \BibitemOpen
\bibfield  {journal} {  }\bibfield  {author} {\bibinfo {author} {\bibfnamefont
  {Y.}~\bibnamefont {Shou}}, \bibinfo {author} {\bibfnamefont {Y.}~\bibnamefont
  {Wang}}, \bibinfo {author} {\bibfnamefont {L.}~\bibnamefont {Miao}}, \bibinfo
  {author} {\bibfnamefont {S.}~\bibnamefont {Chen}},\ and\ \bibinfo {author}
  {\bibfnamefont {H.}~\bibnamefont {Luo}},\ }\bibfield  {title} {\bibinfo
  {title} {Realization of all-optical higher-order spatial differentiators
  based on cascaded operations},\ }\href@noop {} {\bibfield  {journal}
  {\bibinfo  {journal} {Optics Letters}\ }\textbf {\bibinfo {volume} {47}},\
  \bibinfo {pages} {5981} (\bibinfo {year} {2022})}\BibitemShut {NoStop}%
\bibitem [{\citenamefont {Tu}\ \emph {et~al.}(2025)\citenamefont {Tu},
  \citenamefont {Liang}, \citenamefont {Li}, \citenamefont {Xiong},
  \citenamefont {Wu}, \citenamefont {Ren}, \citenamefont {Liu},\ and\
  \citenamefont {Liu}}]{tu2025inverse}%
  \BibitemOpen
  \bibfield  {author} {\bibinfo {author} {\bibfnamefont {Y.}~\bibnamefont
  {Tu}}, \bibinfo {author} {\bibfnamefont {Y.}~\bibnamefont {Liang}}, \bibinfo
  {author} {\bibfnamefont {R.}~\bibnamefont {Li}}, \bibinfo {author}
  {\bibfnamefont {Z.}~\bibnamefont {Xiong}}, \bibinfo {author} {\bibfnamefont
  {H.}~\bibnamefont {Wu}}, \bibinfo {author} {\bibfnamefont {Y.}~\bibnamefont
  {Ren}}, \bibinfo {author} {\bibfnamefont {Z.}~\bibnamefont {Liu}},\ and\
  \bibinfo {author} {\bibfnamefont {T.}~\bibnamefont {Liu}},\ }\bibfield
  {title} {\bibinfo {title} {Inverse design of pancharatnam-berry phase optical
  elements for all-optical multiple-order spatial differentiation},\
  }\href@noop {} {\bibfield  {journal} {\bibinfo  {journal} {Optics \& Laser
  Technology}\ }\textbf {\bibinfo {volume} {183}},\ \bibinfo {pages} {112314}
  (\bibinfo {year} {2025})}\BibitemShut {NoStop}%
\bibitem [{\citenamefont {Long}\ \emph {et~al.}(2021)\citenamefont {Long},
  \citenamefont {Guo}, \citenamefont {Wang},\ and\ \citenamefont
  {Fan}}]{long2021isotropic}%
  \BibitemOpen
  \bibfield  {author} {\bibinfo {author} {\bibfnamefont {O.~Y.}\ \bibnamefont
  {Long}}, \bibinfo {author} {\bibfnamefont {C.}~\bibnamefont {Guo}}, \bibinfo
  {author} {\bibfnamefont {H.}~\bibnamefont {Wang}},\ and\ \bibinfo {author}
  {\bibfnamefont {S.}~\bibnamefont {Fan}},\ }\bibfield  {title} {\bibinfo
  {title} {Isotropic topological second-order spatial differentiator operating
  in transmission mode},\ }\href@noop {} {\bibfield  {journal} {\bibinfo
  {journal} {Optics Letters}\ }\textbf {\bibinfo {volume} {46}},\ \bibinfo
  {pages} {3247} (\bibinfo {year} {2021})}\BibitemShut {NoStop}%
\bibitem [{\citenamefont {Yariv}\ and\ \citenamefont
  {Yeh}(1983)}]{yariv1983optical}%
  \BibitemOpen
  \bibfield  {author} {\bibinfo {author} {\bibfnamefont {A.}~\bibnamefont
  {Yariv}}\ and\ \bibinfo {author} {\bibfnamefont {P.}~\bibnamefont {Yeh}},\
  }\bibfield  {title} {\bibinfo {title} {Optical waves in crystal propagation
  and control of laser radiation},\ }\href@noop {} {\  (\bibinfo {year}
  {1983})}\BibitemShut {NoStop}%
\bibitem [{\citenamefont {Markos}\ and\ \citenamefont
  {Soukoulis}(2008)}]{markos2008wave}%
  \BibitemOpen
  \bibfield  {author} {\bibinfo {author} {\bibfnamefont {P.}~\bibnamefont
  {Markos}}\ and\ \bibinfo {author} {\bibfnamefont {C.~M.}\ \bibnamefont
  {Soukoulis}},\ }\bibfield  {title} {\bibinfo {title} {Wave propagation: from
  electrons to photonic crystals and left-handed materials},\ }in\ \href@noop
  {} {\emph {\bibinfo {booktitle} {Wave Propagation}}}\ (\bibinfo  {publisher}
  {Princeton University Press},\ \bibinfo {year} {2008})\BibitemShut {NoStop}%
\end{thebibliography}%

\end{document}